\documentclass[twocolumn,english,aps,10pt, notitlepage]{revtex4-1}
\usepackage[T1]{fontenc}
\usepackage[latin9]{inputenc}
\usepackage{color}
\usepackage{amsmath}
\usepackage{graphicx}

\makeatletter
\usepackage{fancyhdr,lipsum}
\usepackage{times}               
\usepackage[table]{xcolor}
\usepackage{tikz}

\usepackage{graphicx}
\usepackage{booktabs}            
\usepackage{array}               
\usepackage{longtable}           
\usepackage{siunitx}

\usepackage{makecell}
\usepackage{tablefootnote}       
\usepackage[colorlinks,linkcolor=blue,anchorcolor=blue,citecolor=blue,urlcolor=blue]{hyperref}

\newcolumntype{P}[1]{>{\hspace{4pt}\raggedright\arraybackslash}p{#1}<{\hspace{4pt}}}

\makeatother

\usepackage{babel}
\begin{document}
\title{Finite-Time Thermodynamics Perspective into Nuclear Power Plant Heat
Cycle}
\author{Fang-Ming Cui}
\address{Graduate School of China Academy of Engineering Physics, No. 10 Xibeiwang
East Road, Haidian District, Beijing, 100193, China}
\author{Hui Dong}
\email{hdong@gscaep.ac.cn}

\address{Graduate School of China Academy of Engineering Physics, No. 10 Xibeiwang
East Road, Haidian District, Beijing, 100193, China}
\date{\today}
\begin{abstract}
Nuclear power plants are prominent examples of heat-to-work conversion
systems, and optimizing their thermodynamic performance offers significant
potential for enhancing energy efficiency. With a development history
of less than a century, optimization trends in nuclear power plants
indicate that classical thermodynamics alone may be insufficient,
particularly when maximizing output power rather than efficiency becomes
the primary focus. This paper re-examines nuclear power plant thermodynamic
cycles through the lens of finite-time thermodynamics, an approach
specifically developed to address the practical requirement of enhancing
power output. Beginning with the simpler Brayton cycle without phase
transitions, we obtain the famous Curzon-Ahlborn formula for efficiency
at maximum power. Subsequently we analyze the more complex Rankine
cycle, which incorporates phase transitions. By explicitly considering
the working fluid undergoing phase transitions within the cycle, we
uncover the inherent trade-off between output power and efficiency.
Additionally, we demonstrate that both the maximum attainable power
and efficiency increase as latent heat rises. These findings shall
provide insights and methodologies for future thermodynamic optimization
of nuclear power plants and other Rankine-type cycle systems.
\end{abstract}
\maketitle

\section{Introduction}

The growing demand for energy has accelerated the search for alternative
sources to replace fossil fuels. Among the various options, nuclear
energy stands out as a viable choice \citep{bodansky2007}. Since
the discovery of nuclear fission \citep{meitner1939}, its potential
to address global energy needs has been clear, prompting rapid efforts
to develop controlled fission technology. The first artificial reactor,
Chicago Pile-1, was born, which initiated the era of human-made self-sustaining
nuclear chain reaction \citep{allardice1949,cochran2009}. And the
nuclear energy transformed into electrical energy was then accomplished
for the first time in a small (1.1 MW(th)) experimental breeder reactor,
EBR-1, in Idaho \citep{cochran2009,vanhaaften1979}. Obninsk Nuclear
Power Plant with a net capacity of around 5 MWe, was firstly connected
to an electricity grid to provide power to residences and businesses
\citep{blokhintsev1974}. The accomplishment of the controlled nuclear
fission technology is an unparalleled step in the development of energy
resources. Nuclear power plants(NPPs), as physical facilities for
the utilization of nuclear energy, are developing vigorously.

Although the commercialization of NPPs made rapid strides in the early
days, high capital and construction costs, insecurity and public opposition
prevented further implement of NPPs \citep{ho2019,smil2000}. After
decades of evolution aiming to enhance the thermal performance and
security, NPPs have been greatly improved \citep{abukhader2009}.
The latest commercial NPPs, such as the CAP1400 and EPR models, have
a net capacity of over 1,000 MWe --- far exceeding the capacity provided
by other energy sources \citep{ieanea2015}. Additionally, the integration
of passive safety systems significantly enhances the safety of these
plants, offering better protection in emergency situations \citep{iaeatech2009}.
As a result, the total thermal capacity of NPPs has resumed growth
in recent years. \citep{iaea2022energy}. The latest World Nuclear
Association statistics indicate that nuclear energy from approximately
440 power reactors now supplies about 9 \% of global electricity.
In 2023, NPPs supplied 2,602 TWh of electricity, accounting for about
one-quarter of the world\textquoteright s low-carbon electricity \citep{wna2024}.

However, the cost of nuclear generating technology is still higher
than traditional fossil fuels and mainstream renewable energy like
solar energy and wind energy \citep{ieanea2015} such that the nuclear
renaissance is still questionable \citep{sovacool2010}. Thus, exploring
new nuclear power plant designs remains an urgent task, with cost-effectiveness
being one of the key areas for optimization \citep{goldberg2011,abukhader2009}.
Although to reduce the proportion of cost efficiency needs to take
into account multiple aspects, power and efficiency are the crucial
factors with theoretical research significance among many parameters
\citep{rothwell2018}.

For the efficiency, Sadi Carnot discovered in 1824 the famous Carnot
efficiency \citep{carnot1824} $\eta_{C}=1-T_{c}/T_{h}$, which is
the maximum efficiency for heat engines working between the two thermal
baths with temperature $T_{c}$ and $T_{h}$. Yet, the Carnot efficiency
is obtained from the quasi-static process with zero power, thus has
limit guidance for designing actual industrial facilities like NPPs,
where the power is the major concern. To bridge the gap between theory
and practice, finite-time thermodynamics were developed. And the efficiency
at maximum power output (EMP) was obtained \citep{yvon1955,novikov1958}
\begin{equation}
\eta_{\textrm{EMP}}=1-\sqrt{T_{c}/T_{h}},
\end{equation}
during their researches for NPPs. This result, later known as $\eta_{\textrm{CA}}=1-\sqrt{T_{c}/T_{h}}$,
was rediscovered by Curzon and Ahlborn from a purely theoretical perspective.
They developed a general endoreversible Carnot cycle model that explicitly
incorporates the irreversibility due to heat transfer between the
heat reservoirs and the working substance during the two quasi-isothermal
processes \citep{curzon1975}. The theory of finite-time thermodynamics
is then established with theoretical considerations apart from NPPs.
Using this theoretical framework, many researchers calculated the
EMP of other finite-time thermodynamic cycles using this theoretical
framework \citep{devos1985,leff1987}. The fact that power and efficiency
of a heat engine cannot reach the maximum simultaneously is theoretically
recognized. The boundary of EMP is then considered from the fundamental
theory of non-equilibrium thermodynamics \citep{vandenbroeck2005,esposito2009},
and obtained in low dissipation limit \citep{esposito2010} as
\begin{equation}
\eta_{C}/2\leq\eta_{\textrm{EMP}}\leq\eta_{C}/(2-\eta_{C}).\label{eq:}
\end{equation}
A universal constraint has been established \citep{long2016efficiency,ryabov2016maximum,shiraishi2016universal,mayuhan2018universal},
which can be identified as an important property to evaluate the thermal
performance of thermodynamic cycles at varying output levels. Besides
the theoretical advancement, the fundamental assumption of $1/\tau$
entropy generation scaling in low dissipation finite-time thermodynamics
was experimentally tested \citep{mayuhan2020} and the power-efficiency
trade-off in a finite-time Carnot cycle was also experimentally tested
\citep{zhai2023}.

These advancements have significantly reshaped methodologies for designing
higher-performance heat engines. With these insights, we return to
NPP design, where the influence of finite-time dynamics on thermodynamic
cycles remains largely unexplored. Moreover, current NPPs employ working
substance that undergoes phase transitions, yet their effects have
not been analyzed within the finite-time thermodynamic framework.
Neglecting these effects in NPP cycles may hamper the optimization
of NPPs. \citep{ahmed2021}. And also the output power of NPPs can
be variable depending on the demand in practical applications. In
this context, the trade-off between power and efficiency is important
to evaluate the thermal performance of thermodynamic cycles at varying
output levels, thereby providing optimal design recommendations for
NPPs operating at different output powers.

\section{nuclear power plant}

Before turning into the finite-time thermodynamics of NPPs, we first
revisit their design and development to highlight potential directions
for thermodynamic improvements.

\subsection{The design of the NPPs}

\begin{figure*}
\centering
\includegraphics[scale=0.35]{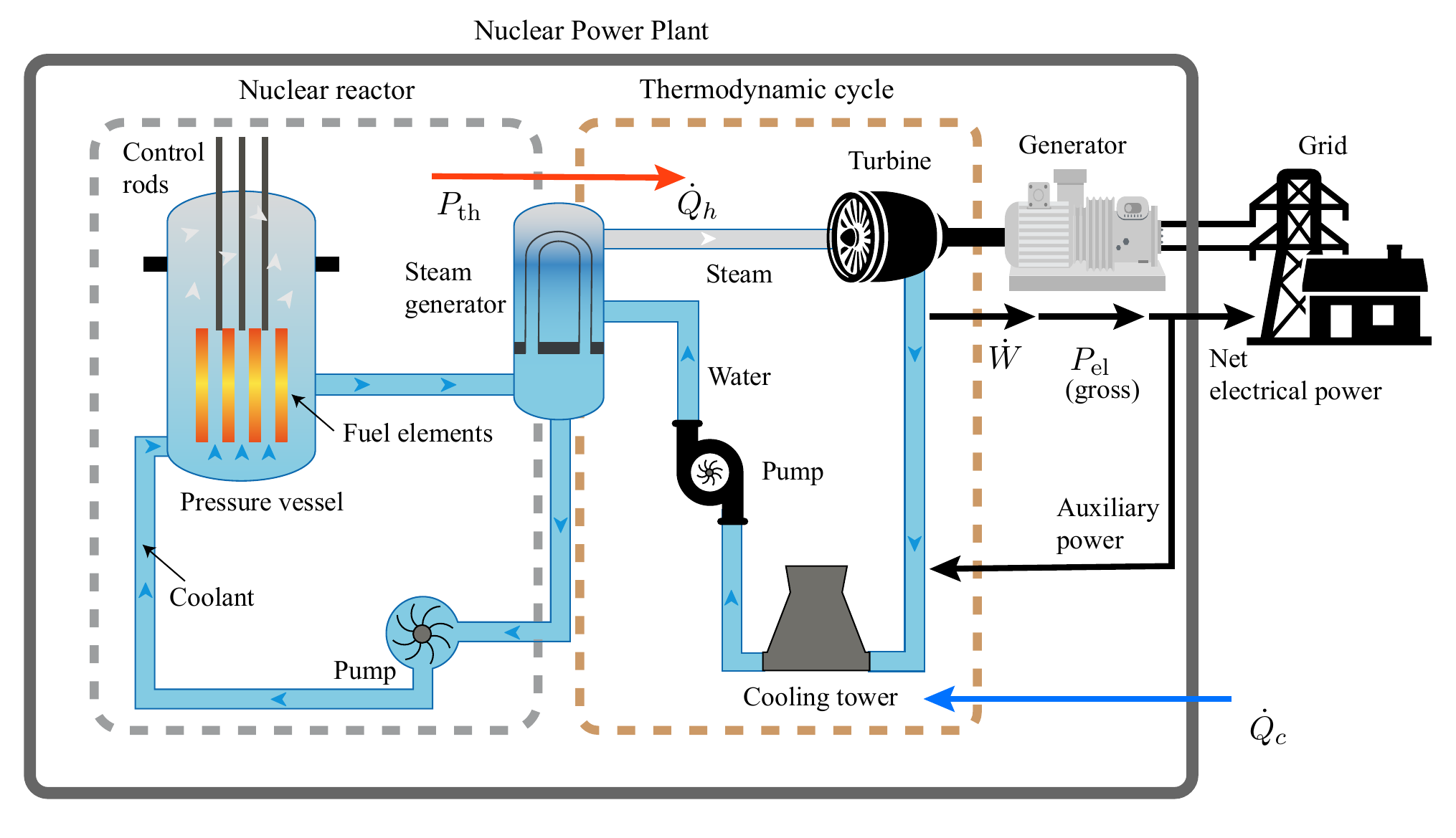}

\caption{\protect\label{fig:Nuclear-Power-Plant}Nuclear power plant structure
and the energy transfer loops. The power generated in the reactor
core is delivered to grid by the two separate closed-loop circuits.
The first loop (gray dotted box) circulates primary coolant through
the reactor core and the steam generator, transferring heat from the
core to the secondary side. The second loop (brown dotted box) carries
the working substance---typically steam---which absorbs heat in
the steam generator and then expands through the turbine to produce
mechanical work. After the turbine, the exhaust steam is condensed
at approximately constant pressure in the condenser to form feedwater.
The feedwater is then pumped back to the steam generator, closing
the thermodynamic cycle (the condenser is cooled by a separate cooling-water
system, often via a cooling tower). $P_{\mathrm{th}}$ is the power
generated by the nuclear reactor. $P_{\mathrm{el}}$ is the output
power of the generator. $\dot{W}$ is the work output by turbine.
$\dot{Q}_{h}$ is the heat power absorbed by the steam generator and
$\dot{Q}_{c}$ is the heat power released to the condenser. Auxiliary
power is generated by the turbine and used inside the steam cycle
such as supplying the pump. Net electrical power is the power delivered
to the electrical grid.}
\end{figure*}

NPPs are capable of converting nuclear energy into electricity. With
the design maturing, the entire process of conversion in a nuclear
power plant is deliberately considered as two parts, illustrated in
Fig. \ref{fig:Nuclear-Power-Plant}.

The first part, illustrated by the gray dotted box in Fig. \ref{fig:Nuclear-Power-Plant},
is responsible for extracting nuclear energy from the nuclear fuel
and converting it into heat. In NPPs, this process takes place within
the nuclear reactor, serving as the primary heat source. This characteristic
distinguishes NPPs from conventional fossil-fuel thermal power plants,
which rely on combustion processes. A nuclear reactor usually consists
of several key components, including the basic fuel (uranium) to generate
heat, a moderator to slow down neutrons, control rods or blades to
regulate or halt the reaction, a coolant to remove or transfer heat,
a steam generator, and a container to transfer heat from the coolant
in the primary circuit to the steam in the secondary circuit\textcolor{red}{{}
}\citep{oka2014}. Based on the properties of the coolant employed,
nuclear reactor configurations can be broadly classified into several
categories---for example, graphite-moderated reactors, light-water
reactors, and heavy-water reactors \citep{wood2007}. For the coolant
with water, there are also different designs depending on whether
the coolant acts as the working substance of the thermodynamic cycle
in the second part, such as the pressurized water reactor (PWR) \citep{cummins2018pwr}
and boiling water reactor (BWR) \citep{fennern2018bwr}.

The second part is to construct the thermodynamic cycle to realize
the heat-work conversion process, as illustrated by the brown dotted
box in Fig. \ref{fig:Nuclear-Power-Plant}. NPPs commonly utilize
a steam-based thermodynamic cycle analogous to conventional fossil-fuel
thermal power plants. Specifically, this cycle comprises four primary
components: the steam generator, the turbine set, the condenser (cooling
tower), and the compressor (pump). This configuration is typically
referred to as the Rankine cycle, the specifics of which will be elaborated
upon subsequently \citep{bowman2020}.

In most of commercial NPPs in operation around the world, each of
the two parts is constructed as a separate closed-loop circuit (PWR
configuration). Energy is transferred between these loops while keeping
their working fluids isolated, ensuring safety and stability \citep{iaea2022nuclear},
shown as Fig. \ref{fig:Nuclear-Power-Plant}. The first loop, commonly
referred to as the reactor loop, involves the circulation through
the reactor core for the coolant, which absorbs heat and achieves
a significantly elevated temperature. Subsequently, the heated coolant
flows into the steam generator, transferring thermal energy across
a pipe wall that separates the coolant from the working substance.
Ultimately, the coolant is recirculated back into the reactor core,
completing the primary loop. In the second loop, depicted by the brown
dotted box in Fig. \ref{fig:Nuclear-Power-Plant}, the working substance,
typically steam, receives heat from the steam generator before entering
the turbine, where mechanical work is produced. After exiting the
turbine, the steam passes through the cooling tower and is condensed
into feedwater under constant pressure. Finally, the feedwater is
pumped back into the steam generator, thereby continuing the thermodynamic
cycle. \citep{bowman2020}.

The output power and efficiency of NPPs are two of the essential design
elements related to the thermodynamic perspective for economic considerations.
Typically, there are two power indicators crucial for one NPP: the
thermal power $P_{\mathrm{th}}$ and the electrical power $P_{\mathrm{el}}$,
which are shown in Fig. \ref{fig:Nuclear-Power-Plant}. Thermal power,
$P_{\mathrm{th}}$, is the rate of heat generation in a nuclear reactor.
In design analyses it is theoretically obtained by integrating the
volumetric power density over the core volume \citep{oka2014}. The
volumetric power density of the nuclear reactor can be determined
by the fuel and the type of the nuclear reactor in the design phase.
And the same type of the nuclear reactors has the approximately same
volumetric power density \citep{oka2014}. During operation, the thermal
power $P_{\mathrm{th}}$ of the nuclear reactor is regulated by adjusting
the position of control rods within the reactor core. Various sensors
positioned upstream and downstream of the steam generator collect
essential parameters, including mass flow rate, pressure, and temperature.
These data are subsequently analyzed by computer systems to calculate
the reactor's current thermal power \citep{wood2003emerging}.

Electrical power $P_{\mathrm{el}}$ is the output power of the electric
generator. In typically NPP reports, the electrical power is further
partitioned into gross electrical power, characterizing the total
output power produced by a NPP before any internal consumption, and
net electrical power, which is the amount of power that is actually
delivered to the electrical grid after subtracting the power used
by the plant itself. The gross electrical power is conceptually closer
to the power that thermodynamics is concerned with so we take $P_{\mathrm{el}}$
to mean gross electrical power throughout. The electrical power $P_{\mathrm{el}}$
can be roughly estimated during the design based on empirical knowledge
about heat transfer and the steam cycle, and is precisely measured
via the output of the electric generator after the construction \citep{iaea2020operating}.

With the definition of the power, the gross efficiency $\eta_{\mathrm{gr}}$
of NPPs is defined to character the performance as \citep{iaea2020operating}
\begin{equation}
\eta_{\mathrm{gr}}=P_{\mathrm{el}}/P_{\mathrm{th}}.
\end{equation}
The efficiency of a typical theoretical calculation of the nuclear
power plant cycle is about $30\%$ lower than the Carnot efficiency
\citep{zohuri2015thermodynamics}. To improve efficiency, standard
measures include raising turbine inlet pressure and temperature and
employing reheat and regenerative feedwater heating. \citep{tanuma2022}.

\subsection{The efficiency improvement}

One of the major purposes of the NPPs design is to improve its efficiency
while keeping the output power. We review how efficiency improvements
have been incorporated across successive NPP generations, as follows.

We have collected key parameters for the four generations of the NPPs,
i.e., Generation I (1957 -1963), Generation II (mid-1960s), Generation
III(1980s) and Generation IV(2020s) \citep{meiswinkel2013}. We have
reviewed data from over 500 nuclear reactors, including those currently
in operation and those planned for construction, to collect information
on coolant outlet temperature $T_{H}$ and the condenser temperature
$T_{C}$ to obtain the corresponding Carnot efficiency $\eta_{\mathrm{C}},$
the thermal power $P_{\mathrm{th}}$ and the electric power $P_{\mathrm{el}}$
to calculate the gross efficiency $\eta_{\mathrm{gr}}$. These data
are obtained from directory of nuclear reactors \citep{iaea1959directory1,iaea1962directory2,iaea1968directory3,iaea1971directory4},
IAEA\textquoteright s Power Reactor Information System (PRIS) \citep{iaea2005thepower,iaea2020operating},
and Handbook of Generation IV Nuclear Reactors \citep{pioro2017handbook}.

\begin{figure}
\begin{centering}
\includegraphics[scale=0.45]{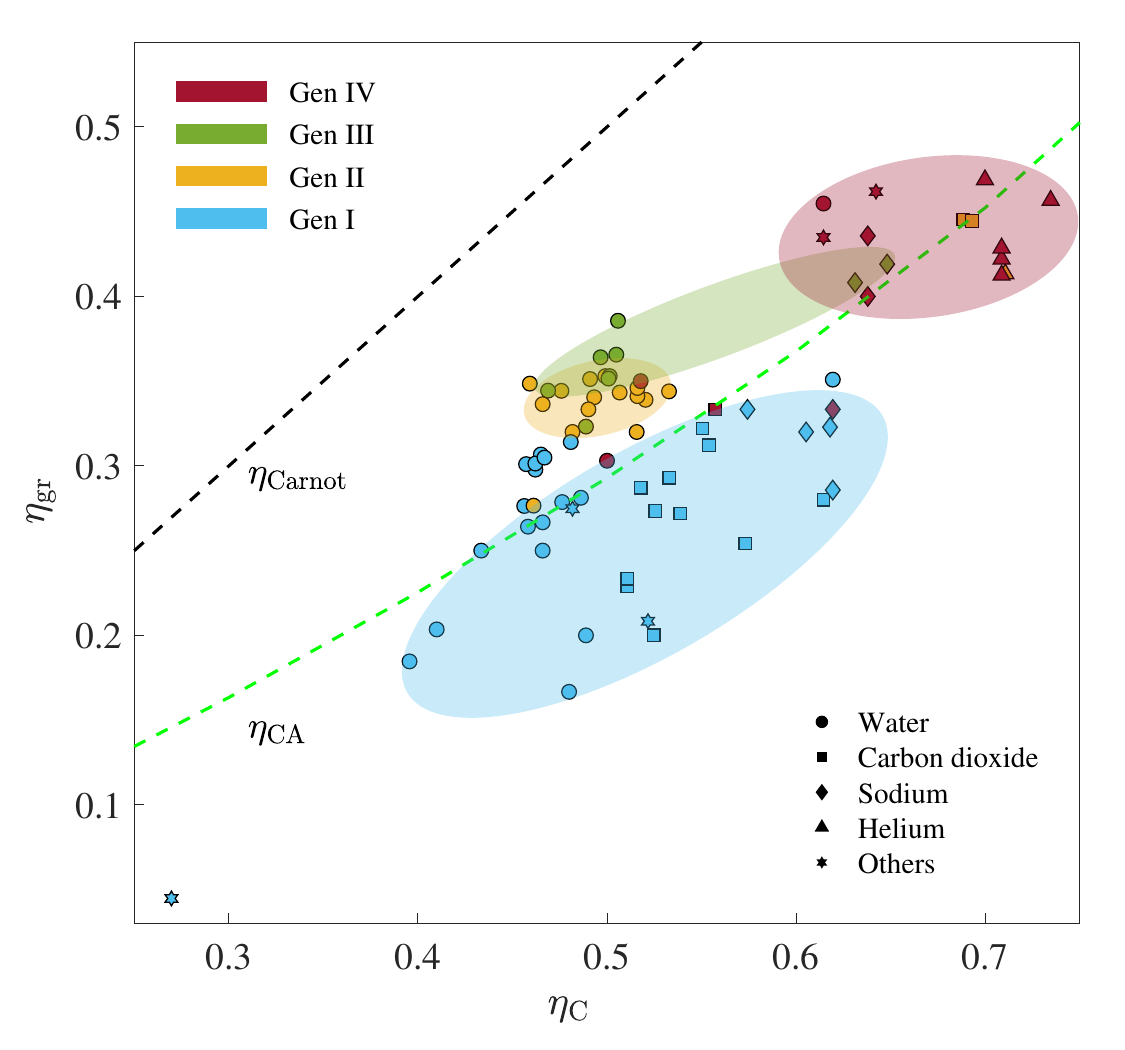}
\par\end{centering}
\caption{\protect\label{fig:GEandCEVarywithGeneration.}Efficiency varies with
generations. Different colors represent the different generations
and different shapes distinguish the coolants applied by these nuclear
power plants. Transparent ellipses are used to show the efficiency
concentration zone of different generations. The black dotted line
represents the Carnot efficiency. The green line represents the CA
efficiency. The data is shown in Appendix A.}
\end{figure}

We present the efficiency improvements for 77 representative NPPs
across different generations in Fig. \ref{fig:GEandCEVarywithGeneration.}.
Each generations is represented by distinct color, and transparent
ellipses are employed to highlight the efficiency concentration areas
clearly. Different coolants are indicated by distinct shapes. The
black dotted line represents the Carnot efficiency, acting as the
upper bound for the actual efficiencies, while the green dotted line
denotes the Curzon-Ahlborn (CA) efficiency, defined as the efficiency
at maximum power according to the finite-time thermodynamic model.
The data for the 77 NPPs is presented in the Appendix A.

As illustrated in Fig. \ref{fig:GEandCEVarywithGeneration.}, gross
efficiency $\eta_{\mathrm{gr}}$ rises in conjunction with Carnot
efficiency. Specifically, the average thermal efficiencies for Generations
I through IV are approximately 26.5\%, 35.2\%, 36.8\%, and 40.7\%,
respectively. The observed improvement in Carnot efficiency is primarily
driven by the increase in the coolant outlet temperature, subsequently
enhancing overall plant efficiency. Notably, a substantial efficiency
enhancement occurs from Generation III to Generation IV, attributed
to the adoption of coolants capable of operating at higher outlet
temperatures $T_{h}$.

Fig. \ref{fig:GEandCEVarywithGeneration.} clearly indicates that
the thermal efficiency of NPPs is proportional to the Carnot efficiency.
And a prevailing method for enhancing thermal efficiency is to elevate
Carnot efficiency by increasing coolant outlet temperatures. However,
the development of the finite-time thermodynamics may bring additional
strategies to the design of NPPs with high efficiency, especially
when power is the primary concern \citep{bejan1996models,ikegami1998}.

Before moving onto the discussion of the NPPs design, we would like
to remark on the difference between the thermodynamic efficiency and
the gross efficiency here. The efficiency of a thermodynamic cycle
defined in finite-time thermodynamics is usually 
\begin{equation}
\begin{aligned}\eta_{\mathrm{th}}=\dot{W}/\dot{Q}_{h},\end{aligned}
\end{equation}
where $\dot{W}$ is the work output per unit time, usually calculated
as $\dot{W}=\dot{Q}_{h}-\dot{Q}_{c}$ under the assumption of no heat
dissipation. $\dot{Q}_{h}$ and $\dot{Q}_{c}$ represent the heat
absorbed or released by the thermodynamic cycle per unit time. From
the point of view of the actual plant, $\dot{W}$ corresponds to the
power output by the turbine, $\dot{Q}_{h}$ corresponds to the heat
power absorbed by the steam generator and $\dot{Q}_{c}$ corresponds
to the heat power released to the condenser. However, the definition
of the gross efficiency for the NPPs has included the efforts to reduce
the heat loss via reheating and regenerative cycles \citep{meitner1939}.
Therefore, the gross efficiency will be used as a reference for comparison,
rather than a criterion, in the later discussion, where the thermodynamic
efficiency $\eta_{\mathrm{th}}$ is primarily considered from the
perspective of finite-time thermodynamics.

\section{The application of finite-time thermodynamics on nuclear power plant}

The thermodynamic cycle in NPPs is idealized as four successive processes:
(i) isobaric expansion, (ii) adiabatic expansion, (iii) isobaric compression,
and (iv) adiabatic compression. The NPPs schematic diagram of the
cycle is illustrated in Fig.\,\ref{fig:Nuclear-Power-Plant}. The
isobaric expansion--expanding at constant pressure-- takes place
in the steam generator, where the working substance absorbs heat in
counter-flow and undergoes a liquid-to-vapor phase transition. The
vapor then expands adiabatically through the turbine, delivering mechanical
work. Upon entering the cooling tower, the vapor-to-liquid phase transition
(condensation) occurs and releases heat at nearly constant pressure..
Finally, the liquid is pressurized adiabatically by the feed-water
pump, restoring the system to its initial state and completing the
cycle. The cycle is typically divided into two categories---Brayton
and Rankine---depending on whether a phase transition occurs.

\subsection{Brayton cycle}

\begin{figure*}
\begin{centering}
\includegraphics[scale=0.8]{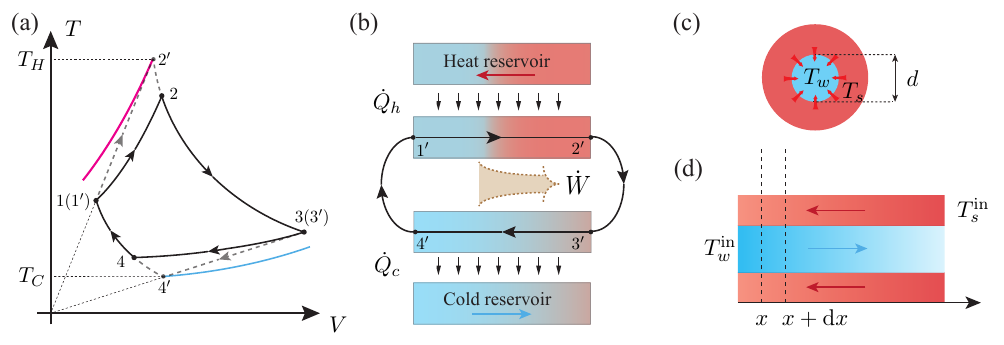}
\par\end{centering}
\caption{\protect\label{fig:Schematic-diagram-ofBC} Schematic diagrams of
the Brayton cycle. (a) Temperature--volume ( $T-V$ ) diagram illustrating
both quasi-static and finite-time Brayton cycles. The light-colored
dashed lines with the endpoints $1^{\prime}\rightarrow2^{\prime}\rightarrow3^{\prime}\rightarrow4^{\prime}$
represents the quasi-static Brayton cycle. Processes $1^{\prime}\rightarrow2^{\prime}$
and $3^{\prime}\rightarrow4^{\prime}$ are quasi-static isobaric expansion
and the contraction processes. Processes $2^{\prime}\rightarrow3^{\prime}$
and $4^{\prime}\rightarrow1^{\prime}$ are the two quasi-static adiabatic
processes. The finite-time Brayton cycle is represented the solid
black lines with the ending points $1\rightarrow2\rightarrow3\rightarrow4$.
The carmine line indicates the temperature changes of the heat reservoir
during the isobaric expansion, while the blue line represents the
temperature changes of the cold reservoir during the isobaric contraction.
(b) Diagram of heat-work conversion in the quasi-static Brayton cycle.
The variations of color represents the changes of the temperature
for both working substance and the reservoirs. (c) Transverse cross-sectional
view of the heat exchanger facilitating the isobaric processes. The
diameter of the heat exchanger is $d$. Temperatures of the heat reservoir
($T_{s}$) and the working substance ($T_{w}$) are assumed homogeneous
within this cross-section. (d) Longitudinal cross-sectional view along
the cylinder axis of the heat exchanger. It is assumed that the fluid
within the heat exchanger reaches steady-state conditions with a stable
temperature gradient. Over incremental distances, heat absorbed by
the working substance equals heat released by the heat reservoir,
with heat transfer rates governed by Newton's cooling law.}
\end{figure*}

For the case without any phase transition, the cycle is known as Brayton
cycle. Before considerting the finite-time thermodynamics, we firstly
analyze the state change of the working substance of a quasi-static
Brayton cycle, shown as the loop enclosed by the black dashed line
with endpoints $1^{\prime}\rightarrow2^{\prime}\rightarrow3^{\prime}\rightarrow4^{\prime}\rightarrow1^{\prime}$
in Fig. \ref{fig:Schematic-diagram-ofBC} (a). The quasi-static isobaric
processes occur in contact with the reservoirs of the highest temperature
$T_{H}$ (hot) and the lowest temperature $T_{C}$ (cold), respectively.

Fig. \ref{fig:Schematic-diagram-ofBC} (b) depicts the cycle of the
heat-work conversion within the quasi-static Brayton cycle. The absorption
and release of heat are accomplished by the heat exchanger. In quasi-static
isobaric expansion (compression), the final temperature of the working
substance can reach the reservoir's highest (lowest) temperature $T_{H}$
($T_{C}$). The efficiency of quasi-static Brayton cycle is determined
by the pressure ratio $P_{H}/P_{C}$ of isobaric expansion and isobaric
compression, and the specific heat ratio $\gamma$ as \citep{bejan2016advanced}
\begin{equation}
\eta_{\mathrm{Br}}=1-(P_{H}/P_{C})^{\frac{\gamma-1}{\gamma}},\label{eq:BraytonCycle=000020efficiency}
\end{equation}
where $P_{H}$ is the pressure of working substance in isobaric expansion
and $P_{C}$ is the pressure in isobaric expansion. Such efficiency
is proved to be lower than the corresponding Carnot efficiency calculated
by the $T_{H}$ and $T_{C}$ \citep{moran2010fundamentals}.

To see the effect of the finite-time thermodynamics, we consider the
dynamics of heat transfer processes during the isobaric expansion
and compression in the finite-time Brayton cycle illustrated as the
black solid lines ($1\rightarrow2\rightarrow3\rightarrow4\rightarrow1$)
in Fig. \ref{fig:Schematic-diagram-ofBC} (a). Due to the insufficient
heat exchange, the temperature of the working substance during the
isobaric expansion (compression) is slightly lower (higher) than that
of the quasi-static ones. To qualitatively consider such difference,
we model the heat exchanger as two coaxial cylinders with Fig. \ref{fig:Schematic-diagram-ofBC}
(c) and (d) showing its transverse and longitudinal cross sections.
Temperatures of the heat reservoir ($T_{s}$) and the working substance
($T_{w}$) are assumed homogeneous on the transverse cross-section,
yet with a distribution along the longitudinal cross section as $T_{s}(x)$
and $T_{w}(x)$.

Heat exchange rate between the reservoir and the working substance
for a small longitudinal distance $\mathrm{d}x$ follows Newton's
cooling law $\dot{Q}=k\mathrm{d}S(T_{s}-T_{w})$, where $k$ is the
coefficient of heat conduction and $\mathrm{d}S=\pi d\mathrm{d}x$
is the area with $d$ as the diameter of the heat exchanger. Such
heat change rate can be rewritten as $\dot{Q}=\dot{m}_{s}c_{ps}\mathrm{d}T_{s}=\dot{m}_{w}c_{pw}\mathrm{d}T_{w}$
from the perspective of the internal energy of between heat resource
and working substance. Here $\dot{m}_{s}(\dot{m}_{w})$ is the mass
flow rate of heat reservoir (working substance) and $c_{ps}(c_{pw})$
is the heat capacity at constant pressure of heat reservoir (working
substance). Assuming no other heat leakage, the amount of heat released
by the heat reservoir per unit length is equal to the amount of heat
absorbed by the working substance as 
\begin{equation}
\begin{aligned}\dot{m}_{s}c_{ps}\frac{\textrm{d}T_{s}}{\textrm{d}x} & =k\pi d(T_{s}-T_{w}),\\
\dot{m}_{w}c_{pw}\frac{\textrm{d}T_{w}}{\textrm{d}x} & =k\pi d(T_{s}-T_{w}).
\end{aligned}
\end{equation}
The temperature distributions are solved with boundary conditions
$T_{s}(x=L)=T_{s}^{\mathrm{in}}$, $T_{w}(x=0)=T_{w}^{\mathrm{in}}$
as follows

\begin{equation}
\begin{aligned}T_{s}= & \frac{(T_{s}^{\mathrm{in}}-T_{w}^{\mathrm{in}})[u_{w}\exp(\mathcal{R}x(\frac{1}{u_{s}}-\frac{1}{u_{w}}))-u_{s}]}{u_{w}\exp(\mathcal{R}L(\frac{1}{u_{s}}-\frac{1}{u_{w}}))-u_{s}}+T_{w}^{\mathrm{in}},\\
T_{w}= & \frac{(T_{s}^{\mathrm{in}}-T_{w}^{\mathrm{in}})[u_{s}\exp(\mathcal{R}x(\frac{1}{u_{s}}-\frac{1}{u_{w}}))-u_{s}]}{u_{w}\exp(\mathcal{R}L(\frac{1}{u_{s}}-\frac{1}{u_{w}}))-u_{s}}+T_{w}^{\mathrm{in}},
\end{aligned}
\label{eq:=000020temp=000020distribution}
\end{equation}
where $u_{w}=\dot{m}_{w}c_{pw}$, $u_{s}=\dot{m}_{s}c_{ps}$ and $\mathcal{R}=\pi kd$.
Here, the superscript $\mathrm{in}$ represents the temperature at
the inlet of heat exchanger. To obtain the corresponding dynamics,
we take the isobaric expansion ($1\rightarrow2$) as an example. The
relation of temperature $T_{1}$ (before the isobaric expansion) and
$T_{2}$ (after the isobaric expansion) is written as 
\begin{equation}
T_{2}=\frac{(T_{H}-T_{1})[u_{s}^{H}\exp(\mathcal{R}_{H}L_{H}(\frac{1}{u_{s}^{H}}-\frac{1}{u_{w}}))-u_{s}^{H}]}{u_{w}\exp(\mathcal{R}_{H}L_{H}(\frac{1}{u_{s}^{H}}-\frac{1}{u_{w}}))-u_{s}^{H}}+T_{1},
\end{equation}
where $L_{H}$ is the length of the heat exchanger. The superscript\,$H$
and subscript\,$H$ denote quantities evaluated while the system is
in contact with the hot reservoir, namely, $u_{s}^{H}=\dot{m}_{s}^{H}c_{ps}^{H}$
and $\mathcal{R}_{H}=\pi k{}_{H}d_{H}$. For simplicity we rewrite
the equation above as 
\begin{equation}
\frac{(u_{s}^{H}-u_{w})E_{H}}{u_{w}E_{H}-u_{s}^{H}}\times T_{1}+T_{2}=\frac{u_{s}^{H}(E_{H}-1)}{u_{w}E_{H}-u_{s}^{H}}\times T_{H},
\end{equation}
where $E_{H}=\exp(\mathcal{R}_{H}L_{H}(\frac{1}{u_{s}^{H}}-\frac{1}{u_{w}}))$.

Similar consideration is applied for the compression isobaric processes
($3\rightarrow4$) in the Brayton cycle with Eq. (\ref{eq:=000020temp=000020distribution}).
We use the superscripts and subscripts $C$ to represent the quantities
are in contact with the cold reservoir. 
\begin{equation}
\begin{aligned}\frac{(u_{s}^{C}-u_{w})E_{C}}{u_{w}E_{C}-u_{s}^{C}}\times T_{3}+T_{4} & =\frac{u_{s}^{C}(E_{C}-1)}{u_{w}E_{C}-u_{s}^{C}}\times T_{C}\end{aligned}
,
\end{equation}
where $E_{C}=\exp(\mathcal{R}_{C}L_{C}(\frac{1}{u_{s}^{C}}-\frac{1}{u_{w}}))$
and the number subscript 3, 4 corresponds to the endpoints of the
isobaric compression process in Fig. \ref{fig:Schematic-diagram-ofBC}(a).
Together with the adiabatic relation $(P_{1}/P_{4})^{\frac{1-\gamma}{\gamma}}T_{1}=T_{4}$
and $(P_{1}/P_{4})^{\frac{1-\gamma}{\gamma}}T_{2}=T_{3}$, we get
exact expressions of the temperature $T_{1}$ and $T_{2}$ as \begin{widetext}
\begin{equation}
{\small {\small {\footnotesize T_{1}=\frac{(u_{w}-u_{s}^{C})u_{s}^{H}(E_{H}-1)T_{H}-\frac{1}{r_{t}}u_{s}^{C}(E_{C}-1)(u_{w}E_{H}-u_{s}^{H})T_{C}}{(u_{s}^{H}-u_{w})(u_{w}-u_{s}^{C})E_{H}-(u_{w}E_{H}-u_{s}^{H})(u_{s}^{C}E_{C}-u_{w})},}}}
\end{equation}

\begin{equation}
{\small {\small {\small {\footnotesize {\footnotesize {\footnotesize T_{2}=\frac{\frac{1}{r_{t}}u_{s}^{C}(E_{C}-1)(u_{s}^{H}-u_{w})E_{H}T_{C}-u_{s}^{H}(E_{H}-1)(u_{s}^{C}E_{C}-u_{w})T_{H}}{(u_{s}^{H}-u_{w})(u_{w}-u_{s}^{C})E_{H}-(u_{w}E_{H}-u_{s}^{H})(u_{s}^{C}E_{C}-u_{w})},}}}}}}
\end{equation}
\end{widetext}where $r_{t}\equiv T_{1}/T_{4}=(P_{1}/P_{4})^{\frac{1-\gamma}{\gamma}}$
is the temperature ratio. Here, we use $P_{1}$ and $P_{4}$ as parameters
($P_{H}=P_{1}$ and $P_{C}=P_{4}$) to represent the pressures at
the end points 1 and 4.

With $\dot{Q}_{h}=\dot{m}_{w}c_{pw}(T_{2}-T_{1})$ and $\dot{Q_{c}}=\dot{m}_{w}c_{pw}(T_{3}-T_{4})$,
we obtain the exact expression for the power $\mathcal{P}=\dot{Q}_{h}-\dot{Q_{c}}$
and the efficiency $\eta=1-\dot{Q_{c}}/\dot{Q}_{h}$, as \begin{widetext}
\begin{equation}
\begin{aligned}\mathcal{P}= & (1-r_{t})(\frac{1}{r_{t}}T_{C}-T_{H})\frac{u_{w}u_{s}^{H}u_{s}^{C}(E_{C}-1)(E_{H}-1)}{(u_{s}^{H}-u_{w})(u_{w}-u_{s}^{C})E_{H}-(u_{w}E_{H}-u_{s}^{H})(u_{s}^{C}E_{C}-u_{w})}\end{aligned}
,\label{eq:Power=000020of=000020Ideal=000020BraytonCycle}
\end{equation}
\end{widetext}
\begin{equation}
\eta=1-r_{t}.
\end{equation}
The power is influenced by many parameters, while the efficiency is
only determined by the temperature ratio $r_{t}$. As the function
of $r_{t}$, the power $\mathcal{P}$ reaches its maximum with the
condition $r_{t}=\sqrt{T_{C}/T_{H}}$ when other parameters, e.g.
$u_{w}$, are fixed. The corresponding efficiency is 
\begin{equation}
\eta=1-\sqrt{\frac{T_{C}}{T_{H}}},\label{eq:15}
\end{equation}
which is the well-known CA efficiency. The result in Eq. (\ref{eq:15})
indicates that the efficiency of maximum power is achieved by adjusting
the pressures $P_{1}$ and $P_{4}$ as $(P_{1}/P_{4})^{\frac{1-\gamma}{\gamma}}=\sqrt{T_{C}/T_{H}}$
in practical application.

\begin{figure}
\centering
\begin{centering}
\includegraphics[scale=0.45]{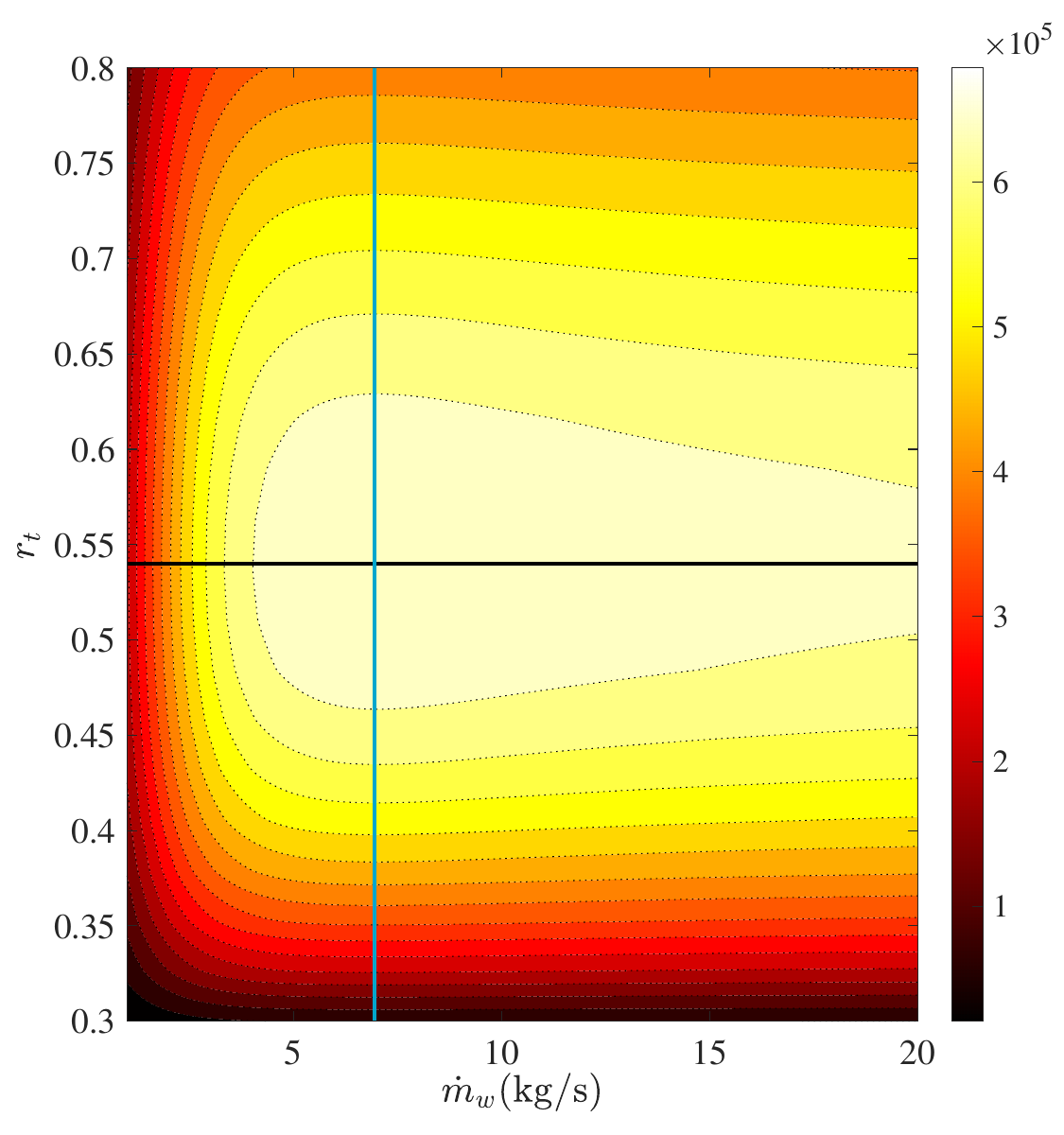}
\par\end{centering}
\caption{\protect\label{fig:PowerandefficiencyofBC} Power as the function
of the temperature ratio $r_{t}$ and mass flow rate $\dot{m}_{w}$.
Parameters used in calculations: hot and cold reservoir temperatures
$T_{H}=1023.15\textrm{K}$ and $T_{C}=298.15\textrm{K}$; coefficients
of heat conduction in hot and cold side $k_{H}=k_{C}=3000\textrm{W}\textrm{m}^{-2}\textrm{K}^{-1}$;
specific heat capacities $c_{pw}=c_{ps}^{H}=c_{ps}^{C}=1.004\times10^{3}\textrm{J}\textrm{kg}^{-1}\textrm{K}^{-1}$;
hot and cold side heat exchanger lengths $L_{H}=L_{C}=2\textrm{m}$;
hot and cold side heat exchanger diameters $d_{H}=d_{C}=0.6\textrm{m}$.
The power attains a maximum when the temperature ratio and flow rate
are optimized independently. The black solid line marks the optimal
value $r_{t}=\sqrt{T_{C}/T_{H}}$.}
\end{figure}

Another parameter affecting the output power is the flow rate $\dot{m}_{w}$
of the working substance. With Eq. (\ref{eq:Power=000020of=000020Ideal=000020BraytonCycle}),
we calculate the power as the function of flow rate $\dot{m}_{w}$
and temperature ratio $r_{t}$ with results shown in Fig. \ref{fig:PowerandefficiencyofBC}.
When the temperature ratio $r_{t}$ is fixed, a maximum power is obtained
at certain mass flow rate, represented by the\textbf{ }cyan\textbf{
}solid line in Fig. \ref{fig:PowerandefficiencyofBC}. We understand
the existence of such optimal flow as follows. The magnitude of the
flow indicates how far the system is from equilibrium. When the mass
flow rate is small, it is close to quasi-static resulting in the small
power. While the mass flow rate is large, the heat exchange between
the working substance and reservoir is insufficient also leading to
the small power. The results indicate that the power of the Brayton
cycle can be tuned by adjusting these parameters, which have optimal
values. This provides a specific guidance for the practical operation
of the Brayton cycle under varying loads. 

\subsection{Rankine cycle}

\begin{figure}
\begin{centering}
\includegraphics[scale=0.7]{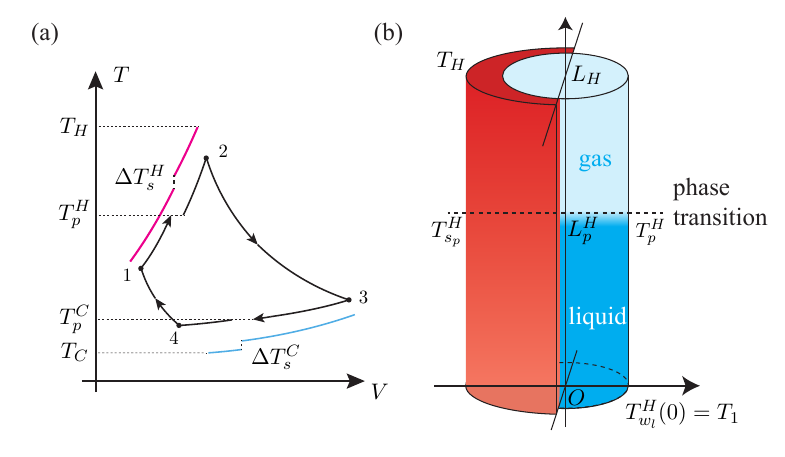}
\par\end{centering}
\caption{\protect\label{fig:Heat-exchanger-model} Heat exchanger model with
phase transition. (a) The $T-V$ diagram of the Rankine cycle. The
black lines with arrows show the Rankine cycle of the working substance.
In the isobaric expansion ($1\rightarrow2$) and isobaric expansion
($3\rightarrow4$ ), the working substance undergoes the phase transition
at the phase transition temperatures $T_{p}^{H}$ and $T_{p}^{C}$.
In phase transition processes, the temperature maintains and the volume
changes which is represented by the gap in the black lines. The carmine
(blue) line represents the temperature changes of the hot (cold) reservoir.
Due to the assumption of phase transition of the working substance
occurring in one position, the temperatures of the reservoirs have
a sudden change. (b) The heat exchanger model with phase transition
in the isobaric expansion process. The outer cylinder in red color
represents the hot reservoir transferring heat with stable temperature
distribution. The inner cylinder represents the working substance
with the navy(wathet) blue part representing the liquid (gas) phase.
$L$ is the total length of the cylinder and $T_{H}$ is the inlet
temperature of hot reservoir. $T_{w_{l}}^{H}(0)=T_{1}$ is the inlet
temperature of the working substance. $L_{p}^{H}$ is the phase transition
position, where the temperature of hot reservoir has a sudden changes
from $T_{s_{p}}^{H}$ to $T_{s_{l}}^{H}.$}
\end{figure}

In this subsection, we study the finite-time Rankine cycle, which
describs the steam cycle in NPP plants. The main difference between
the Rankine cycle and the Brayton cycle is that the Rankine cycle
involves phase transitions---specifically, liquid-vapor transitions---during
the isobaric expansion and isobaric compression processes, whereas
the Brayton cycle operates entirely with a single-phase gas throughout.
Fig. \ref{fig:Heat-exchanger-model} (a) shows the diagram of the
Rankine cycle. The black solid lines represent the finite-time Rankine
cycle. The carmine(blue) line shows the temperature of the hot(cold)
reservoir. During phase transition, the temperature of working substance
remains at the constant phase-transition temperature $T_{p}^{H}$($T_{p}^{C})$
while the hot (cold) reservoir temperature undergoes a sudden step
of magnitude $\Delta T_{s}^{H}$($\Delta T_{s}^{C}$). Fig. \ref{fig:Heat-exchanger-model}
(b) presents a detailed model of the processes occurring inside the
heat exchanger. In this heat exchanger model, the phase transition
is assumed to occur at a certain point $L_{p}^{H}$. 
Each phase of the working substance undergoes the same heat-transfer
process as in the previously discussed Brayton cycle. For every phase
segment, we derive the heat-transfer equations separately for the
liquid and vapor regions and present their formal solutions, using
the same analytical framework applied to the Brayton cycle.

When contacting the hot reservoir, the working substance is firstly
in the liquid state, the temperature distributions of reservoir $T_{s_{l}}^{H}$
and working substance $T_{w_{l}}^{H}$ are 
\begin{equation}
\begin{aligned}T_{s_{l}}^{H}= & \frac{u_{l}}{u_{s}^{H}}C_{2}\exp(\mathcal{R}_{l}^{H}x(\frac{1}{u_{s}^{H}}-\frac{1}{u_{l}}))+C_{1},\\
T_{w_{l}}^{H}= & C_{2}\exp(\mathcal{R}_{l}^{H}x(\frac{1}{u_{s}^{H}}-\frac{1}{u_{l}}))+C_{1},
\end{aligned}
\label{eq:liquidHotHE}
\end{equation}
where $k_{l}$ is the coefficient of heat conduction in the liquid
segment; $u_{l}=\dot{m}_{w_{l}}c_{pw_{l}}$ and $u_{s}^{H}=\dot{m}_{s}^{H}c_{ps}^{H}$;
$\mathcal{R}_{l}^{H}=\pi k_{l}d_{H}$. $C_{i},\,i\in\{1,2\}$ are
the coefficients determined by boundary conditions $T_{w_{l}}^{H}(0)=T_{1}$
and $T_{w_{l}}^{H}(L_{p}^{H})=T_{p}^{H}$. And the temperature distributions
in gaseous segment has the similar form
\begin{equation}
\begin{aligned}T_{s_{g}}^{H}= & \frac{u_{g}}{u_{s}^{H}}C_{4}\exp(\mathcal{R}_{g}^{H}x(\frac{1}{u_{s}^{H}}-\frac{1}{u_{g}}))+C_{3},\\
T_{w_{g}}^{H}= & C_{4}\exp(\mathcal{R}_{g}^{H}x(\frac{1}{u_{s}^{H}}-\frac{1}{u_{g}}))+C_{3},
\end{aligned}
\label{eq:gasHotHe}
\end{equation}
where $k_{g}$ is the coefficient of heat conduction in the gas segment;
$u_{g}=\dot{m}_{w_{g}}c_{pw_{g}}$; $\mathcal{R}_{g}^{H}=\pi k_{g}d_{H}$;
$C_{i},\,i\in\{3,4\}$ are the coefficients determined by boundary
conditions $T_{s_{g}}^{H}(L_{H})=T_{H}$ and $T_{w_{g}}^{H}(L_{p}^{H})=T_{p}^{H}$.

At the phase transition position, the temperature of working substance
is the phase transition temperature determined by the property of
the working substance. The temperature of hot reservoir undergoes
a sudden change satisfying 
\begin{equation}
\begin{aligned}u_{s}^{H}(T_{s_{g}}^{H}(L_{p}^{H})-T_{s_{l}}^{H}(L_{p}^{H}) & )=\dot{m}_{w}\lambda_{H}\end{aligned}
,\label{eq:BoundaryCondition}
\end{equation}
where $\lambda_{H}$ is the latent heat of phase change for the working
substance. For simplicity, we define two parameters $T_{s_{p}}^{H}=T_{s_{g}}^{H}(L_{p}^{H})$
and $T_{s_{l}}^{H}(L_{p}^{H})=T_{s_{p}}^{H}+\frac{\dot{m}_{w}\lambda_{H}}{u_{s}^{H}}$.

Given the hot reservoir temperature $T_{H}$, the working substance\textquoteright s
inlet temperature $T_{1}$, and the phase-transition temperature $T_{p}^{H}$
at the operating pressure $P_{1}$, the outlet temperature after isobaric
heat addition, $T_{2}$, is obtained by firstly solving the nonlinear
system derived above (see Eq. (\ref{eq:HETempDistrFunc})) and then
substituting $x=L_{H}$ into Eq. (\ref{eq:gasHotHe})

\begin{equation}
\begin{aligned}T_{1}= & C_{2}+C_{1},\\
T_{p}^{H}= & C_{2}\exp(\mathcal{R}_{l}^{H}L_{p}^{H}(\frac{1}{u_{s}^{H}}-\frac{1}{u_{l}}))+C_{1},\\
T_{p}^{H}= & C_{4}\exp(\mathcal{R}_{g}^{H}L_{p}^{H}(\frac{1}{u_{s}^{H}}-\frac{1}{u_{g}}))+C_{3},\\
T_{s_{p}}^{H}= & \frac{u_{l}}{u_{s}^{H}}C_{2}\exp(\mathcal{R}_{l}^{H}L_{p}^{H}(\frac{1}{u_{s}^{H}}-\frac{1}{u_{l}}))+C_{1},\\
T_{s_{p}}^{H}= & \frac{u_{g}}{u_{s}^{H}}C_{4}\exp(\mathcal{R}_{g}^{H}L_{p}^{H}(\frac{1}{u_{s}^{H}}-\frac{1}{u_{g}}))+C_{3}-\frac{\dot{m}_{w}\lambda_{H}}{u_{s}^{H}},\\
T_{H}= & \frac{u_{g}}{u_{s}^{H}}C_{4}\exp(\mathcal{R}_{g}^{H}L_{H}(\frac{1}{u_{s}^{H}}-\frac{1}{u_{g}}))+C_{3}.
\end{aligned}
\label{eq:HETempDistrFunc}
\end{equation}
This establishes a relation between the inlet temperature $T_{1}$
and the outlet temperature $T_{2}$ of the working substance. For
simplicity, we denote this relationship as
\begin{equation}
T_{2}=\mathcal{H}(T_{1}).
\end{equation}
After the same analysis, the equations contacting with the cold reservoir
in Rankine cycle also is acquired. Also we denote the relationship
as 
\begin{equation}
T_{4}=\mathcal{L}(T_{3}).
\end{equation}

The adiabatic function of liquid is as follows
\begin{equation}
\begin{aligned}[][\kappa_{T}c_{v}+\alpha^{2}T\exp(\alpha T-\kappa_{T}P+\mathcal{C})]\textrm{d}T\\
-\kappa_{T}\alpha T\exp(\alpha T-\kappa_{T}P+\mathcal{C})\textrm{d}P & =0
\end{aligned}
,\label{eq:liquid=000020adiabatic=000020equation-1}
\end{equation}
where the coefficient of thermal expansion $\alpha$ and the isothermal
compressibility $\kappa_{T}$ is assumed to be constant. $\mathcal{C}(T_{0},P_{0})$
is a constant, which depends on the initial condition for solving
the above equation. Here we take the standard condition with $T_{0}=300\mathrm{K}$
and $P_{0}=10^{5}\mathrm{Pa}$.

Furthermore, the dependence of the latent heat $\lambda_{H}(P)$ on
the pressure must be considered. In our calculations, we incorporate
real data for saturated water and saturated vapor to impose this dependence.
Fig. \ref{fig:function-of-saturated} in the Appendix C illustrates
the relationship between latent heat and the corresponding phase-transition
temperature.

Each subprocesses are connected to form the whole Rankine cycle by
the following functions.

\paragraph*{1-2 isobaric expansion process}

\begin{equation}
T_{2}=\mathcal{H}(T_{1}),\,P_{1}=P_{2}.\label{eq:RankineEQ1}
\end{equation}

\paragraph*{2-3 adiabatic expansion process}

\begin{equation}
r_{t}T_{2}=T_{3}.\label{eq:RankineEQ2}
\end{equation}

\paragraph*{3-4 isobaric compression process}

\begin{equation}
T_{4}=\mathcal{L}(T_{3}),\,P_{3}=P_{4}.\label{eq:RankineEQ3}
\end{equation}

\paragraph*{4-1 adiabatic compression process}

\begin{equation}
\begin{aligned}[][\kappa_{T}c_{v}+\alpha^{2}T\exp(\alpha T-\kappa_{T}P+\mathcal{C})]\textrm{d}T\\
-\kappa_{T}\alpha T\exp(\alpha T-\kappa_{T}P+\mathcal{C})\textrm{d}P & =0
\end{aligned}
.\label{eq:RankineEQ4}
\end{equation}

Equations (\ref{eq:RankineEQ1}), (\ref{eq:RankineEQ2}), (\ref{eq:RankineEQ3})
and (\ref{eq:RankineEQ4}) determine the temperatures $T_{i}\,(i\in\{1,2,3,4\})$
of working substance at process turning points of the Rankine cycle.
After the design parameters, e.g., the dimensions of the heat exchanger,
are given, the power and efficiency is obtained as the function of
these temperatures 
\begin{equation}
\begin{aligned}\mathcal{P}= & [(c_{pl}-c_{pg})(T_{s_{p}}^{H}-T_{s_{p}}^{C})+c_{pl}(T_{4}-T_{1})\\
 & -c_{pg}(T_{3}-T_{2})+(\lambda_{H}-\lambda_{C})]\dot{m}_{w}
\end{aligned}
,\label{eq:PowerofRankineCycle}
\end{equation}
\begin{equation}
\begin{aligned}\eta= & 1-\frac{c_{pg}(T_{3}-T_{s_{p}}^{C})+c_{pl}(T_{s_{p}}^{C}-T_{4})+\lambda_{C}}{c_{pl}(T_{s_{p}}^{H}-T_{1})+c_{pg}(T_{2}-T_{s_{p}}^{H})+\lambda_{H}}\end{aligned}
.\label{eq:EffofRankineCycle}
\end{equation}

To illustrate the dependence of power and efficiency on system parameters,
we vary the mass flow rate $\dot{m}_{w}$ and high-temperature pressure
$P_{1}$ within specified ranges. We use high-temperature gas-cooled
reactors(HTR-PM)\citep{kugeler2018} as an example to compare the
numerical results with the actual situation. The parameters introduced
in the derivation---such as geometric dimensions and heat-transfer
coefficients---match the actual design specifications of the HTR-PM
reactor \citep{kugeler2018}. The full set of these parameters is
presented in the Appendix D. In the numerical model, we fit the phase-transition
temperature $T_{p}(P)$ and latent heat $\lambda_{H}(P)$ for water
as a function of the pressure with details in the Appendix C.

\begin{figure}
\begin{centering}
\includegraphics[scale=0.45]{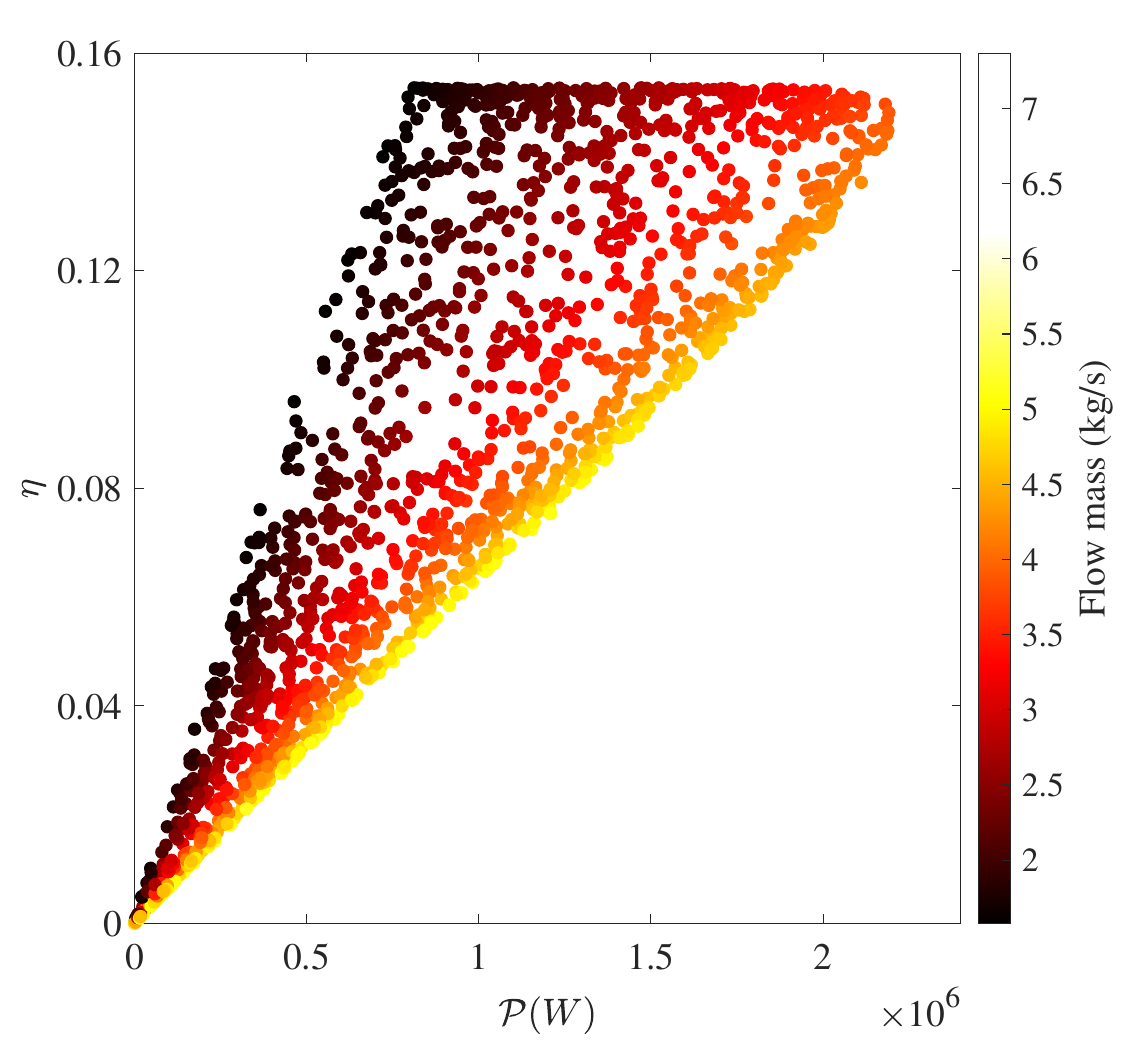}
\par\end{centering}
\caption{\protect\label{fig:PandEfficonstrainofFTRC}Power and efficiency constraint
of the finite time Rankine cycle. We use HTR-PM nuclear power plant
as a reference for parameter selection. Parameters setting is shown
in the Appendix D. Each dot in the diagram is calculated by the different
mass flow rate $\dot{m}_{w}$ and high-temperature pressure $P_{1}$.
Dot color indicates the magnitude of the mass flow rate $\dot{m}_{w}$.
$P_{1}$ ranges from $5\textrm{MPa}$ to $20\textrm{MPa}$ and $\dot{m}_{w}$
ranges from $1.58\textrm{kg}/\textrm{s}$ to $7.89\textrm{kg}/\textrm{s}$.
The range of $P_{1}$ and $\dot{m}_{w}$ resembles the real situation
applied in HTR-PM nuclear power plant. }
\end{figure}

The power and efficiency, as Eq. (\ref{eq:PowerofRankineCycle}) and
(\ref{eq:EffofRankineCycle}) shown, depend on many parameters which
impact on the temperature $T_{i}$ ($i=1,2,3,4$) of Rankine cycle.
For practical purposes, we focus on two operationally adjustable variables---the
mass flow rate $\dot{m}_{w}$ and the high-temperature pressure $P_{1}$.
We sample $\dot{m}_{w}$ and $P_{1}$ over prescribed ranges and compute
the corresponding power and efficiency for each sample. The resulting
graphical representations of power and efficiency are presented in
Fig. \ref{fig:PandEfficonstrainofFTRC}. The color of each dot encodes
the mass flow rate $\dot{m}_{w}$. We have excluded the cases where
the solution yields negative output power. In Fig. \ref{fig:PandEfficonstrainofFTRC},
it is clear to see the constraint between the power and efficiency,
and the efficiency is improved when the flow decreases. The power
has the maximum as the mass flow rate $\dot{m}_{w}$ and high-temperature
pressure $P_{1}$ varying. Our results indicate that the level of
output power can be increased by adjusting the parameters of NPPs,
e.g. the flow rate.

We would like to mention that the real thermal power and efficiency
of HTR-PM are around $250\mathrm{MW}$ and 40\%. We choose parameters
to match  maximum power with the power for HTR-PM, and then calculate
the efficiency. Under the same power, our theory predicts a relative
lower efficiency. This is because the efficiency of nuclear power
plants is improved by the inclusion of regenerative cycles.

With the purpose of searching alternative working substance to improve
thermodynamic cycle performance, we find the substance\textquoteright s
latent heat emerges as a crucial parameter governing the power and
efficiency of thermodynamic cycle, particularly when the cycle includes
a phase-change stage such as NPPs. To examine the impact of the phase
transition on a Rankine-type cycle, we vary the latent heat $\lambda_{H}$---the
latent heat associated with the liquid-to-vapour transformation of
water---and recompute the net power output and efficiency. Latent
heat is modified by adjusting the constant term in the empirical correlation
derived from experimental data. 

\begin{figure*}
\begin{centering}
\includegraphics[scale=0.24]{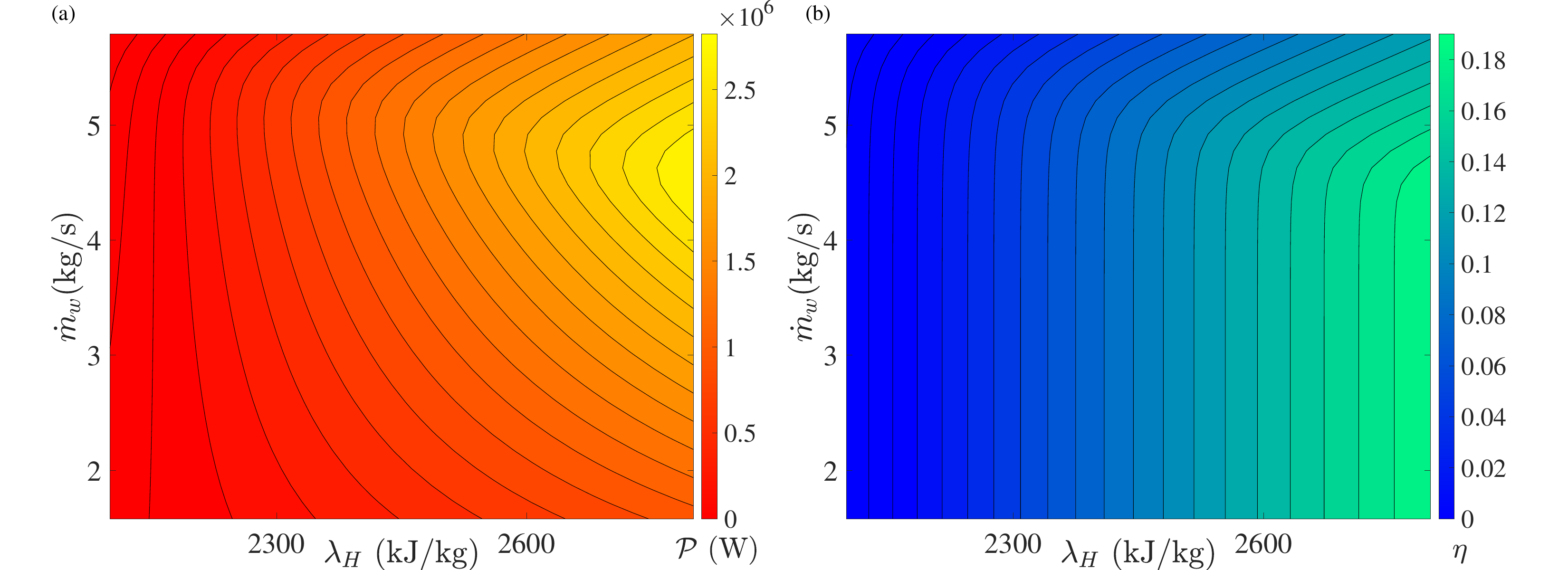}
\par\end{centering}
\begin{centering}
\includegraphics[scale=0.24]{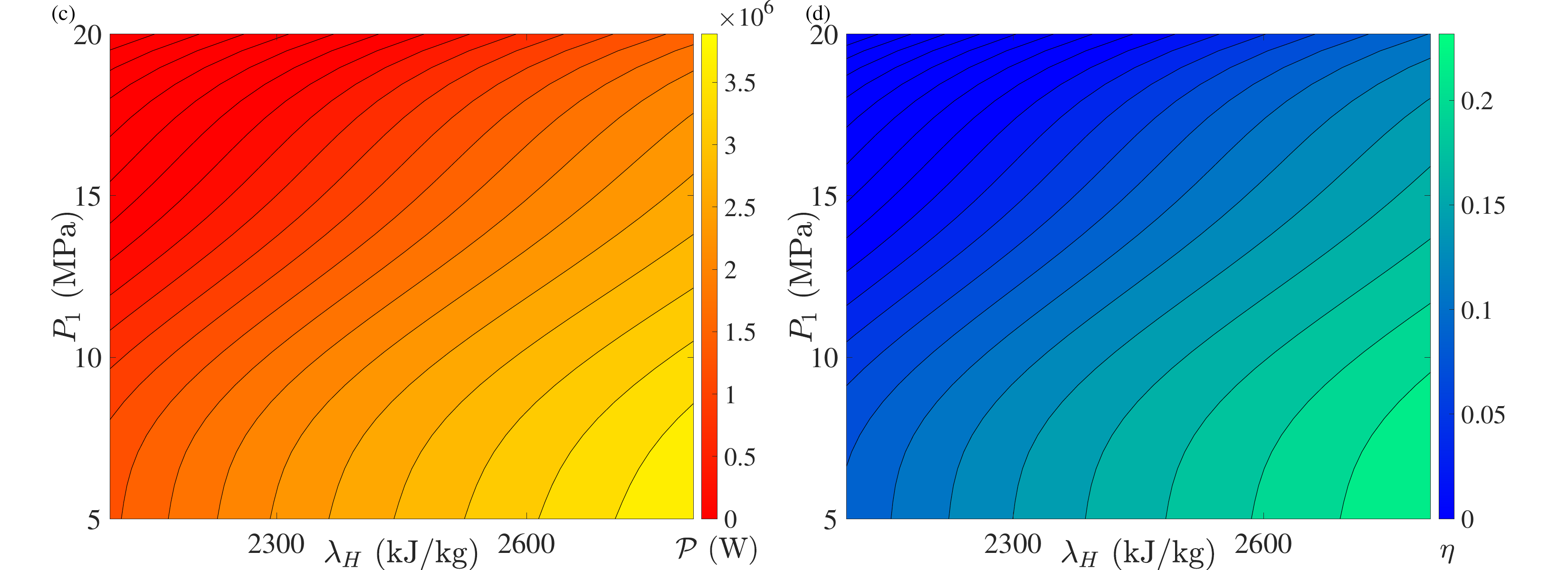}
\par\end{centering}
\caption{\protect\label{fig:PandEfficiencywithDifferentLatent}Power and efficiency
are the functions of the latent heat $\lambda_{H}$, the high-temperature
pressure $P_{1}$ and the mass flow rate $\text{\ensuremath{\dot{m}}}_{w}$
of the work substance. (a) Power as a function of the mass flow rate
$\dot{m}_{w}$ and latent heat $\lambda_{H}$ with a fixed pressure
value $P_{1}=15\textrm{MPa}$. (b) Efficiency as a function of the
mass flow rate $\dot{m}_{w}$ and latent heat $\lambda_{H}$ with
a fixed pressure value $P_{1}=15\textrm{MPa}$. (c) Power as a function
of the high-temperature pressure $P_{1}$ and latent heat $\lambda_{H}$
with a fixed flow mass value $\dot{m}_{w}=5\textrm{kg}/\textrm{s}$.
(d) Efficiency as a function of the high-temperature pressure $P_{1}$
and latent heat $\lambda_{H}$ with a fixed flow mass value $\dot{m}_{w}=5\textrm{kg}/\textrm{s}$.}
\end{figure*}

Fig. \ref{fig:PandEfficiencywithDifferentLatent}(a) and (b) are the
power and efficiency contour as the function of mass flow rate $\dot{m}_{w}$
and latent heat $\lambda_{H}$ with a fixed high-temperature pressure
$P_{1}=15\textrm{MPa}$. And Fig. \ref{fig:PandEfficiencywithDifferentLatent}(c)
and (d) are the power and efficiency contour as the function of high-temperature
pressure $P_{1}$ and latent heat $\lambda_{H}$ with a fixed mass
flow rate $\dot{m}_{w}=5\textrm{kg}/\textrm{s}$. As shown in Fig.
\ref{fig:PandEfficiencywithDifferentLatent}(a-d), the power and efficiency
increases with the latent heat increasing. In our model, the increase
in latent heat of phase change enhances the capacity of heat absorption
in the cycle, thereby increasing the power. In Fig. \ref{fig:PandEfficiencywithDifferentLatent}(a),
it shows that, for fixed latent heat $\lambda_{H}$, the power exhibits
a maximum for particular mass flow rate $\dot{m}_{w}$. And as $\lambda_{H}$
increasing, the power is monotonically increasing in the range which
conforms to our prediction. Fig. \ref{fig:PandEfficiencywithDifferentLatent}(b)
reports the efficiency as a function of the mass flow rate $\dot{m}_{w}$
and latent heat $\lambda_{H}$. Larger latent heat $\lambda_{H}$
yields higher efficiency, whereas increasing mass flow rate $\dot{m}_{w}$
reduces efficiency because the operation departs from the quasi-static
(reversible) regime. In Fig. \ref{fig:PandEfficiencywithDifferentLatent}(c)
and (d), the power and efficiency both monotonic increase as the high-temperature
pressure $P_{1}$ increases for fixed $\lambda_{H}$. The higher $P_{1}$
means the higher outlet temperature $T_{2}$ in adiabatic expansion
resulting in more heat absorption thus increases the power and efficiency.
And when $P_{1}$ is fixed, the power and efficiency increase as $\lambda_{H}$
increase. This result implies that when choosing working substances,
it is necessary to select those with large latent heat of phase transition.

\section{Conclusion}

In practical nuclear power plants, thermal conversion involves a complex
process that balances considerations of both safety and thermal efficiency.
We investigate the their historical development of the structure of
NPPs and summarize optimization strategies across four generations
of nuclear power plants through investigating the crucial data of
a NPP from the first to fourth generation plants. The collected data
indicates that the prevailing method to enhance efficiency involves
increasing the coolant inlet temperature, thereby elevating the theoretical
upper bound given by the Carnot efficiency.

However, the practical application of Carnot efficiency includes considerable
redundancy, imposes stringent requirements on coolant selection and
structural materials, and offers limited practical guidance during
actual load variations. Here, we offer the new perspective through
the lens of finite-time thermodynamics by re-analyzing a simplified
thermodynamic cycles representative of nuclear power plants. From
the Brayton cycle without phase transitions, we established the model
and extended to the analysis to Rankine cycles that involve phase
transitions. With the model, we calculated their respective power
outputs and efficiencies and showed alignment of the results with
the established results for the efficiency at the maximum power efficiency
for the Brayton cycle.

For the Rankine cycle, we explicitly included the phase transition
into the model, and reveal that variations in the mass flow rate $\dot{m}_{w}$
and high-temperature pressure $P_{H}$ significantly influence both
power and efficiency, with power exhibiting an identifiable maximum
under specific conditions. Utilizing different working substance,
each characterized by unique phase-transition constraint equations,
yields markedly different constraints on achievable power and efficiency.
Specifically, increases in latent heat during high-temperature phase
transitions directly raise the maximum attainable efficiency. Our
work particularly emphasizes how augmenting latent heat influences
the Rankine cycle, demonstrating through calculations that elevated
latent heat enhances both power output and efficiency. For the future
nuclear power plant design by changing the working substance, these
findings shall offer new theoretical insights.

\section{Appendix}

\subsection{The Evolution of Nuclear Power Plant}

We provide here the data sets of the 77 nuclear power plants used
in the main context. And the data are categorized as four generations
\citep{meiswinkel2013} in chronological order. Table I lists representative
nuclear power plants (NPPs) and their thermal parameters for Generations
I--IV, compiled from International Atomic Energy Agency (IAEA) sources.
\onecolumngrid
\begingroup
\small
\setlength{\tabcolsep}{1.5pt}
\setlength{\LTleft}{\fill}
\setlength{\LTright}{\fill}

\begin{longtable}{@{}%
  >{\raggedright\arraybackslash}p{.27\textwidth}
  S[table-format=1.0,  table-column-width=.055\textwidth]
  >{\raggedright\arraybackslash}p{.105\textwidth}
  S[table-format=4.1,  table-column-width=.095\textwidth]
  S[table-format=3.3,  table-column-width=.095\textwidth]
  S[table-format=3.2,  table-column-width=.110\textwidth]
  S[table-format=2.2,  table-column-width=.105\textwidth]
  S[table-format=2.2,  table-column-width=.105\textwidth]
@{}}
\caption{Historical nuclear power plants (selected).}\label{tab:npp}\\
\toprule
\rowcolor{gray!20}
\multicolumn{1}{l}{\textbf{Name}} &
\multicolumn{1}{c}{\textbf{Gen.}} &
\multicolumn{1}{l}{\textbf{Coolant}} &
\multicolumn{1}{c}{\shortstack{\textbf{Thermal}\\\textbf{power (MW)}}} &
\multicolumn{1}{c}{\shortstack{\textbf{Electric}\\\textbf{power (MW)}}} &
\multicolumn{1}{c}{\shortstack{\textbf{Coolant outlet}\\\textbf{temperature}\\\textbf{(\si{\celsius})}}} &
\multicolumn{1}{c}{\shortstack{\textbf{Condenser}\\\textbf{temperature}\\\textbf{(\si{\celsius})}}} &
\multicolumn{1}{c}{\shortstack{\textbf{Carnot}\\\textbf{efficiency (\%)}}}
\\
\midrule
\endfirsthead

\toprule
\rowcolor{gray!20}
\multicolumn{1}{l}{\textbf{Name}} &
\multicolumn{1}{c}{\textbf{Gen.}} &
\multicolumn{1}{l}{\textbf{Coolant}} &
\multicolumn{1}{c}{\shortstack{\textbf{Thermal}\\\textbf{power (MW)}}} &
\multicolumn{1}{c}{\shortstack{\textbf{Electric}\\\textbf{power (MW)}}} &
\multicolumn{1}{c}{\shortstack{\textbf{Coolant outlet}\\\textbf{temperature}\\\textbf{(\si{\celsius})}}} &
\multicolumn{1}{c}{\shortstack{\textbf{Condenser}\\\textbf{temperature}\\\textbf{(\si{\celsius})}}} &
\multicolumn{1}{c}{\shortstack{\textbf{Carnot}\\\textbf{efficiency (\%)}}}
\\
\midrule
\endhead

\midrule
\multicolumn{8}{r}{\emph{Continued on next page}}\\
\endfoot

\bottomrule
\endlastfoot

\rowcolors{2}{white}{gray!04}
Obninsk Nuclear Power Plant (First Atomic Power Plant of the USSR) & 1 & light water & 30 & 5 & 300.00 & 25.00 & 47.99 \\
Belgian Thermal Reactor BR-3 & 1 & light water & 40.9 & 11.5 & 307.00 & 25.00 & 48.62 \\
Consolidated Edison Thorium Reactor & 1 & light water & 585 & 163 & 296.00 & 25.00 & 47.63 \\
Shippingport Atomic Power Station & 1 & light water & 225 & 67 & 281.00 & 25.00 & 46.21 \\
Stationary Medium Power Plant-1 & 1 & light water & 10 & 2.035 & 232.00 & 25.00 & 40.99 \\
Voronezh Atomic Power Station & 1 & light water & 760 & 210 & 275.00 & 25.00 & 45.62 \\
Yankee Atomic Electric Company & 1 & light water & 392 & 118 & 276.00 & 25.00 & 45.72 \\
Dresden Nuclear Power Station & 1 & light water & 626 & 192 & 284.00 & 25.00 & 46.50 \\
Experimental Boiling Water Reactor (EBWR) & 1 & light water & 20 & 5 & 253.00 & 25.00 & 43.35 \\
Elk River Reactor & 1 & light water & 73 & 22 & 281.00 & 25.00 & 46.21 \\
Kahl Experimental Power Station (Main) & 1 & light water & 60 & 16 & 285.00 & 25.00 & 46.59 \\
Pacific Gas \& Electric Plant & 1 & light water & 164 & 50 & 286.00 & 25.00 & 46.69 \\
Ulyanovsk Atomic Power Station & 1 & light water & 250 & 50 & 310.00 & 25.00 & 48.89 \\
Ural Atomic Power Station & 1 & light water & 285 & 100 & 510.00 & 25.00 & 61.94 \\
Vallecitos Boiling Water Reactor & 1 & light water & 50 & 12.5 & 285.00 & 25.00 & 46.59 \\
Carolinas-Virginia Tube Reactor (CVTR) & 1 & heavy water & 60.5 & 19 & 301.00 & 25.00 & 48.08 \\
Nuclear Power Demonstration Station & 1 & heavy water & 83.3 & 22 & 277.00 & 25.00 & 45.82 \\
Swedish Reactor R-3/ADAM & 1 & heavy water & 65 & 12 & 220.00 & 25.00 & 39.55 \\
Advanced Gas-cooled Reactor & 1 & carbon dioxide & 100 & 28 & 500.00 & 25.00 & 61.45 \\
Berkeley Nuclear Power Station & 1 & carbon dioxide & 557.4 & 160 & 345.00 & 25.00 & 51.78 \\
Bradwell Nuclear Power Station & 1 & carbon dioxide & 531 & 171 & 390.00 & 25.00 & 55.05 \\
Calder Hall Reactor & 1 & carbon dioxide & 201 & 46 & 336.00 & 25.00 & 51.07 \\
Chapelcross Reactors & 1 & carbon dioxide & 180 & 42 & 336.00 & 25.00 & 51.07 \\
Centrale de Chinon EDF-1 & 1 & carbon dioxide & 300 & 82 & 355.00 & 25.00 & 52.55 \\
Centrale de Chinon EDF-2 & 1 & carbon dioxide & 785 & 230 & 365.00 & 25.00 & 53.29 \\
Reactor G2 (G3) & 1 & carbon dioxide & 200 & 40 & 354.00 & 25.00 & 52.47 \\
Hinkley Point Nuclear Power Station & 1 & carbon dioxide & 960 & 261 & 373.00 & 25.00 & 53.87 \\
Hunterston Nuclear Generating Station & 1 & carbon dioxide & 535 & 167 & 395.00 & 25.00 & 55.39 \\
Heavy Water Moderated Gas-Cooled Power Reactor of Czechoslovakia & 1 & carbon dioxide & 590 & 150 & 425.00 & 25.00 & 57.31 \\
Experimental Breeder Reactor II (EBR-II) & 1 & sodium & 62.5 & 20 & 482.00 & 25.00 & 60.53 \\
Enrico Fermi Atomic Power Plant & 1 & sodium & 300 & 100 & 427.00 & 25.00 & 57.43 \\
Hallam Nuclear Power Facility & 1 & sodium & 254 & 82 & 507.00 & 25.00 & 61.79 \\
Sodium Reactor Experiment & 1 & sodium & 21 & 6 & 510.00 & 25.00 & 61.94 \\
Reactor G1 & 1 & air & 38 & 1.7 & 135.00 & 25.00 & 26.96 \\
Piqua Organic Moderated Reactor & 1 & organic & 45.5 & 12.5 & 302.00 & 25.00 & 48.17 \\
Dounreay Fast Reactor & 1 & sodium-potassium & 72 & 15 & 350.00 & 25.00 & 52.17 \\
Kerncentrale Doel, Unit 1 & 2 & light water & 1192 & 410 & 337.00 & 12.00 & 53.28 \\
Kozloduy Nuclear Power Station, Units 1 and 2 & 2 & light water & 1375 & 440 & 302.00 & 25.00 & 48.17 \\
RWE Biblis Nuclear Power Station, Block A & 2 & light water & 3540 & 1200 & 316.00 & 9.50 & 52.04 \\
Ohi Nuclear Power Station, Units 1 and 2 & 2 & light water & 3423 & 1175 & 325.00 & 22.00 & 50.67 \\
Novovoronezh Nuclear Power Station, Unit 2 & 2 & light water & 1320 & 365 & 280.00 & 25.00 & 46.11 \\
Novovoronezh Nuclear Power Station, Units 3 and 4 & 2 & light water & 1375 & 440 & 301.00 & 5.00 & 51.57 \\
Kewaunee Nuclear Power Plant & 2 & light water & 1721 & 586 & 315.00 & 25.00 & 49.32 \\
Three Mile Island Nuclear Station, Unit 1 & 2 & light water & 2568 & 876 & 318.00 & 13.00 & 51.61 \\
Three Mile Island Nuclear Station, Unit 2 & 2 & light water & 2772 & 959 & 318.00 & 13.00 & 51.61 \\
Gemeinschaftskernkraftwerk Tullnerfeld & 2 & light water & 2100 & 723 & 286.00 & 20.00 & 47.58 \\
Kernkraftwerk Brunsbuettel & 2 & light water & 2292 & 805 & 285.00 & 11.00 & 49.10 \\
Shimane Nuclear Power Station, Unit 1 & 2 & light water & 1380 & 460 & 286.00 & 12.00 & 49.02 \\
Oskarshamnsverket, Unit 2 & 2 & light water & 1700 & 600 & 286.00 & 7.00 & 49.91 \\
Cooper Nuclear Station & 2 & light water & 2381 & 801 & 285.00 & 25.00 & 46.59 \\
James A. FitzPatrick Nuclear Power Plant & 2 & light water & 2436 & 849 & 278.00 & 25.00 & 45.92 \\
Hinkley Point B Power Station & 2 & carbon dioxide & 1493 & 665 & 645.00 & 13.00 & 68.85 \\
Hunterston B Generating Station & 2 & carbon dioxide & 1496 & 665 & 649.00 & 10.00 & 69.31 \\
THTR Prototype (Kernkraftwerk Uentrop) & 2 & helium & 750 & 310 & 750.00 & 23.00 & 71.07 \\
Advanced Passive PWR (AP1000) & 3 & light water & 3400 & 1200 & 324.70 & 25.00 & 50.14 \\
APR-1400 & 3 & light water & 3983 & 1400 & 323.90 & 25.00 & 50.08 \\
The Evolutionary Power Reactor (EPR) & 3 & light water & 4590 & 1770 & 330.00 & 25.00 & 50.58 \\
VVER-1200 (V-491) & 3 & light water & 3200 & 1170 & 328.90 & 25.00 & 50.49 \\
Advanced Boiling Water Reactor (ABWR) & 3 & light water & 3920 & 1350 & 288.00 & 25.00 & 46.88 \\
Advanced CANDU Reactor 1000 & 3 & light water & 3200 & 1165 & 319.00 & 25.00 & 49.66 \\
Indian 700 MWe PHWR & 3 & light water & 2166 & 700 & 310.00 & 25.00 & 48.89 \\
BN-800 & 3 & sodium & 2100 & 880 & 574.00 & 25.00 & 64.82 \\
BN-600 & 3 & sodium & 1470 & 600 & 535.00 & 25.00 & 63.12 \\
High-Performance Light-Water Reactor (HP-LWR) & 4 & light water & 2300 & 1046 & 500.00 & 25.00 & 61.45 \\
Integrated Modular Water Reactor (IMR) & 4 & light water & 1000 & 350 & 345.00 & 25.00 & 51.78 \\
System-Integrated Modular Advanced Reactor (SMART) & 4 & light water & 330 & 100 & 323.00 & 25.00 & 50.00 \\
KAMADO FBR (CRIEPI, Japan) & 4 & carbon dioxide & 3000 & 1000 & 400.00 & 25.00 & 55.72 \\
BN-1200 & 4 & sodium & 2800 & 1220 & 550.00 & 25.00 & 63.79 \\
4S (Toshiba Energy Systems \& Solutions Corp., Japan) & 4 & sodium & 30 & 10 & 510.00 & 25.00 & 61.94 \\
China Fast Reactor 600 (CFR-600) & 4 & sodium & 1500 & 600 & 550.00 & 25.00 & 63.79 \\
High-Temperature Gas-Cooled Reactor (Pebble-Bed Module) & 4 & helium & 500 & 211 & 750.00 & 25.00 & 70.87 \\
Gas-Turbine High-Temperature Reactor & 4 & helium & 600 & 274 & 850.00 & 25.00 & 73.46 \\
Pebble Bed Modular Reactor (PBMR) & 4 & helium & 400 & 165 & 750.00 & 25.00 & 70.87 \\
General Atomics Prismatic Modular High-Temperature Gas-Cooled Reactor & 4 & helium & 350 & 150 & 750.00 & 25.00 & 70.87 \\
Steam-Cycle High-Temperature Gas-Cooled Reactor & 4 & helium & 625 & 293 & 750.00 & 34.00 & 69.99 \\
Chinese Supercritical Water-Cooled Reactor (CSR-1000) & 4 & supercritical light water & 2300 & 1000 & 500.00 & 25.00 & 61.45 \\
Japanese Supercritical Water-Cooled Reactor (JSCWR) & 4 & supercritical light water & 3681 & 1700 & 560.00 & 25.00 & 64.23 \\

\end{longtable}

\endgroup
\twocolumngrid

\subsection{The liquid adiabatic equation}

The Rankine cycle contains the adiabatic compression with the working
substance in the liquid state. To describe the adiabatic compression,
we give the adiabatic function of liquid. Firstly, we give the liquid
state function in first order approximation. Since the volume $V$
of the liquid usually does not change much, we perform a Taylor expansion
on $\textrm{ln}V$ at the normal condition $(T_{0}=300\mathrm{K},P_{0}=10^{5}\mathrm{Pa})$
and retain it only to the first order to obtain
\begin{equation}
\begin{aligned}\ln V= & \ln V(T_{0},P_{0})+\frac{1}{V}\frac{\partial V}{\partial T}|_{T_{0},P_{0}}(T-T_{0})\\
 & +\frac{1}{V}\frac{\partial V}{\partial P}|_{T_{0},P_{0}}(P-P_{0})
\end{aligned}
.
\end{equation}
Rewrite it as 
\begin{equation}
\begin{aligned}\ln V= & \alpha_{0}T-\kappa_{T_{0}}P+\mathcal{C}\end{aligned}
,
\end{equation}
where $\alpha_{0}=\frac{1}{V}\frac{\partial V}{\partial T}|_{T_{0},P_{0}}$
is the coefficient of thermal expansion at $(T_{0},P_{0})$, $\kappa_{T_{0}}=\frac{1}{V}\frac{\partial V}{\partial P}|_{T_{0},P_{0}}$
is the isothermal compressibility coefficient and $\mathcal{C}=\ln V(T_{0},P_{0})-\alpha_{0}T_{0}+\kappa_{T_{0}}P_{0}=-6.8378$
is the coefficient depends on the initial condition. Assuming $\text{\ensuremath{\alpha_{0}}},\kappa_{T_{0}}$
as constants \citep{smith2018introCET}, we obtain the equation of
state of liquid water as
\begin{equation}
V=\exp(\alpha_{0}T-\kappa_{T_{0}}P+\mathcal{C}).\label{eq:liquid=000020state=000020function}
\end{equation}
With the internal energy function of liquid in differential form
\begin{equation}
\textrm{d}U=(c_{p}-\alpha_{0}PV)\textrm{d}T+(\kappa_{T_{0}}P-\alpha_{0}T)V\textrm{d}P,\label{eq:internal=000020energy=000020function=000020of=000020liquid}
\end{equation}
and the adiabatic condition 
\begin{equation}
\textrm{d}Q=\textrm{d}U+P\textrm{d}V=0,\label{eq:adiabatic=000020condition}
\end{equation}
we obtain the liquid adiabatic equation is as
\begin{equation}
\begin{aligned}[][\kappa_{T_{0}}c_{v}+\alpha_{0}^{2}T\exp(\alpha_{0}T-\kappa_{T_{0}}P+\mathcal{C})]\textrm{d}T\\
-\kappa_{T_{0}}\alpha_{0}T\exp(\alpha T-\kappa_{T_{0}}P+\mathcal{C})\textrm{d}P & =0
\end{aligned}
.\label{eq:liquid=000020adiabatic=000020equation}
\end{equation}

\subsection{Latent heat and phase transition temperature function}

\begin{figure}
\begin{centering}
\includegraphics[scale=0.4]{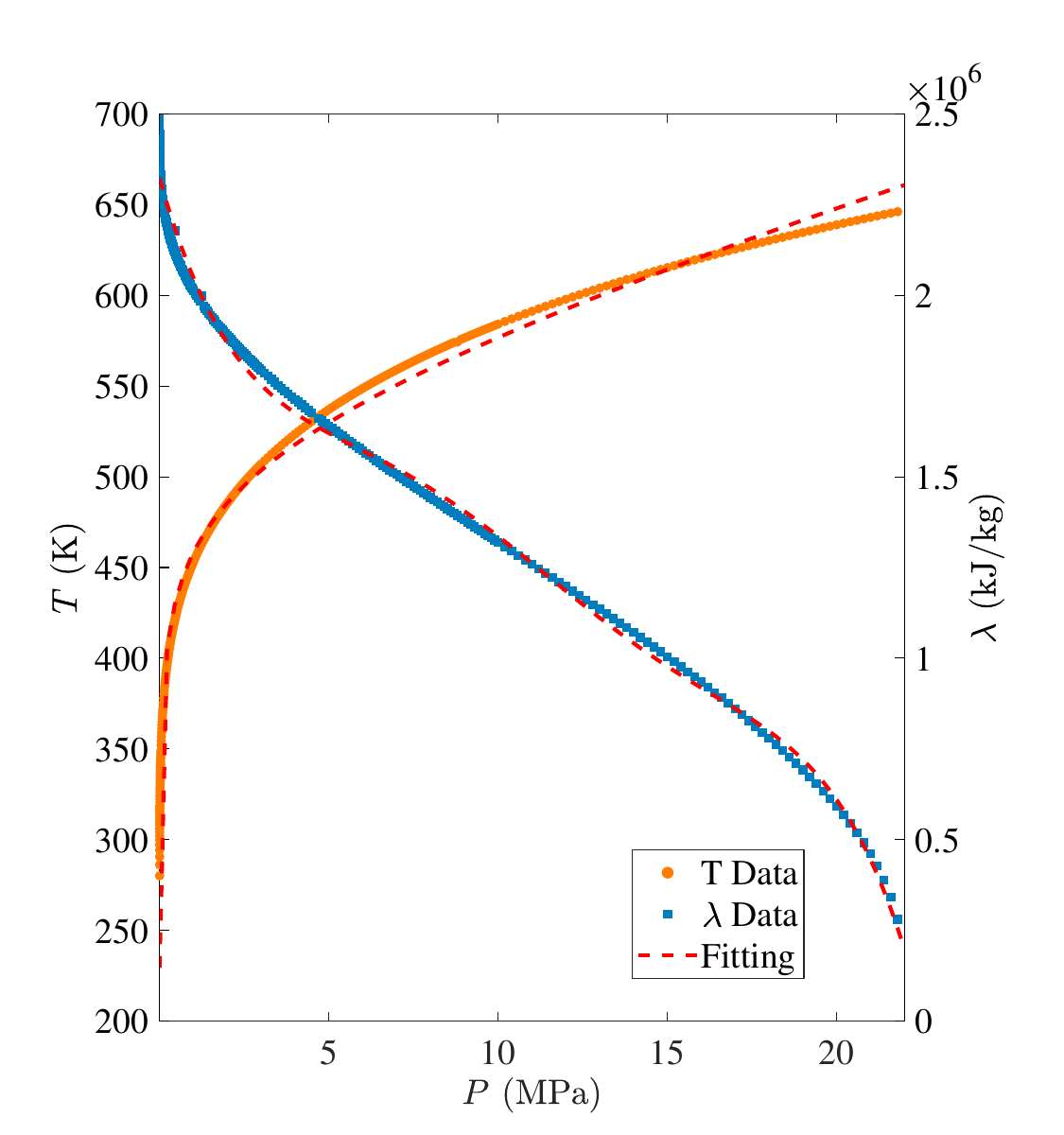}
\par\end{centering}
\caption{\protect\label{fig:function-of-saturated} Actual data and fitted
relationships between latent heat $\lambda$ and pressure $P$, as
well as between phase transition temperature $T$ and pressure $P$.
The red circle(blue square) markers represent actual phase transition
temperature(latent heat) at different pressure, while the red dashed
lines indicate fitted curves.}
\end{figure}

To align the theoretical model more closely with practical conditions,
we fit the empirical data \citep{harvey2014asme} for saturated water
vapor to establish relationship between latent heat and pressure,
as well as the relationship between phase transition temperature and
pressure. Fig. \ref{fig:function-of-saturated} shows the empirical
latent heat $\lambda$ and phase transition temperature $T$ as the
function of pressure. The blue(red) markers are the latent heat $\lambda$(phase
transition temperature $T$) data; the solid lines are the fits. The
$T-P$ fit uses $T(P)=a\mathrm{ln}(b\times P)+c\times P+d$, where
$a,\,b,\,c,\,d$ are coefficients, and the $\lambda-P$ relation is
a quintic polynomial $\lambda(P)=a_{5}P^{5}+a_{4}P^{4}+a_{3}P^{3}+a_{2}P^{2}+a_{1}P+a_{0}$.

\subsection{The parameters selected in the calculation program}

We choose the parameters in our calculation program to reflect the
situation of the real HTR-PM nuclear power plant with details shown
in Table II. For these exact parameters are not available in literature,
we use theoretical or reasonable estimates.

\begin{table}[t]
\centering
\begingroup
\small
\setlength{\tabcolsep}{0pt}      
\rowcolors{2}{white}{gray!06}

\caption{Parameters used in the calculation program versus reported values for \textsc{HTR-PM}.}
\label{tab:htrpm}
\begin{tabular}{@{} P{.24\columnwidth} P{.38\columnwidth} P{.38\columnwidth} @{}}
\toprule
\rowcolor{gray!20}
\textbf{Parameter} & \textbf{simulation} & \textbf{real situation} \\
\midrule
\(T_H\)                  & \SI{1023}{\kelvin}                                 & \SI{1023}{\kelvin} \\
\(T_C\)                  & \SI{298}{\kelvin}                                  & \SI{298}{\kelvin} \\
\(\dot m_w\)             & \SIrange{30}{100}{\kilogram\per\second}            & \SI{99.4}{\kilogram\per\second} \\
\(c_{p\ell}\)            & \SI{4.2}{\kilo\joule\per\kilogram\per\kelvin}      & \SI{4.2}{\kilo\joule\per\kilogram\per\kelvin} \\
\(c_{pg}\)               & \SI{1.5}{\kilo\joule\per\kilogram\per\kelvin}      & \textit{unknown} \\
\(P_{1}\,(P_H)\)         & \SIrange{2}{19}{\mega\pascal}                      & \SI{16}{\mega\pascal} \\
\(\dot m_H\)             & \SI{96.2}{\kilogram\per\second}                    & \SI{96.2}{\kilogram\per\second} \\
\(c_{pH}\)               & \SI{5.19}{\kilo\joule\per\kilogram\per\kelvin}     & \SI{5.19}{\kilo\joule\per\kilogram\per\kelvin} \\
\(\dot m_C\)             & \SI{15000}{\kilogram\per\second}                   & \(\sim\)\SI{15000}{\kilogram\per\second} \\
\(c_{pC}\)               & \SI{4.2}{\kilo\joule\per\kilogram\per\kelvin}      & \SI{4.2}{\kilo\joule\per\kilogram\per\kelvin} \\
\(T_1\)                  & \(\sim\)\SI{350}{\kelvin}                          & \SI{478}{\kelvin} \\
\(T_2\)                  & \(\sim\)\SI{1010}{\kelvin}                         & \SI{839}{\kelvin} \\
\(k_\ell\)               & \SI{3000}{\watt\per\metre\squared\per\kelvin}      & \SI{3000}{\watt\per\metre\squared\per\kelvin} \\
\(k_g\)                  & \SI{3000}{\watt\per\metre\squared\per\kelvin}      & \textit{unknown} \\
\(D_H\)                  & \SI{0.6}{\metre}                                   & \SI{0.58}{\metre} \\
\(D_C\)                  & \SI{5}{\metre}                                     & \textit{unknown} \\
\(L_H\)                  & \SI{8}{\metre}                                     & \SI{8}{\metre} \\
\(L_C\)                  & \SI{1000}{\metre}                                  & \textit{unknown} \\
\(\gamma\)               & \num{1.4}                                          & \textit{unknown} \\
\(\lambda_C\)            & \SI{2428.2}{\kilo\joule\per\kilogram}              & \SI{2428.2}{\kilo\joule\per\kilogram} \\
\(T_C^{P}\)              & \SI{303}{\kelvin}                                  & \SI{303}{\kelvin} \\
\(P_C\)                  & \SI{0.005}{\mega\pascal}                           & \SI{0.005}{\mega\pascal} \\
\(P_{\mathrm{th}}\)      & \(\sim\)\SI{200}{\mega\watt}                       & \SI{250}{\mega\watt} \\
\(P\)                    & \SIrange{20}{60}{\mega\watt}                       & \SI{108}{\mega\watt} \\
\(\eta\)                 & \numrange{0.1}{0.3}                                & \num{0.4} \\
\bottomrule
\end{tabular}

\medskip
\emph{Notes:} "\(\sim\)" denotes approximate values; "unknown" indicates no reported value.
\par\endgroup
\end{table}

\bibliographystyle{apsrev4-2}
\bibliography{NPPRef}

\end{document}